\documentclass[journal=jpclcd,manuscript=article]{achemso}
\usepackage[version=3]{mhchem} 
\usepackage{braket}
\usepackage{subcaption}
\usepackage{xcolor}
\usepackage{amsmath}
\usepackage{siunitx}
\usepackage{booktabs}
\usepackage{graphicx}
\usepackage{makecell}

\NewDocumentCommand{\stau}{}{%
  \mathord{\text{\resizebox{!}{1ex}{\usefont{U}{psy}{m}{n}\symbol{"74}}}}%
}
\usepackage{newunicodechar}
\newunicodechar{π}{$\mathtt{\pi}$}

\author{Brieuc Le D\'e}
\affiliation
{Sorbonne Universit\'e, CNRS, Institut des NanoSciences de Paris, 4 place Jussieu, 75005 Paris, France}%
\altaffiliation{Department of Chemistry and Biochemistry, University of California, Merced CA 95343, USA}
\author{Simon Huppert}
\affiliation[INSP]
{Sorbonne Universit\'e, CNRS, Institut des NanoSciences de Paris, 4 place Jussieu, 75005 Paris, France}%
\author{Riccardo Spezia}
\affiliation[LCT]
{Sorbonne Universit\'e, CNRS, Laboratoire de Chimie Th\'eorique, 4 place Jussieu, 75005 Paris, France}%
\author{Alex W. Chin}
\email{alex.chin@insp.jussieu.fr}
\affiliation[INSP]
{Sorbonne Universit\'e, CNRS, Institut des NanoSciences de Paris, 4 place Jussieu, 75005 Paris, France}%

\title[Vib. Control SBF]
{Vibrational Activation Triggers Ultrafast Excited State Intramolecular Proton Transfer in Single-Benzene Fluorophores}

\abbreviations{SBF}

\begin{document}

\begin{tocentry}

      \includegraphics[width=1.0\linewidth]{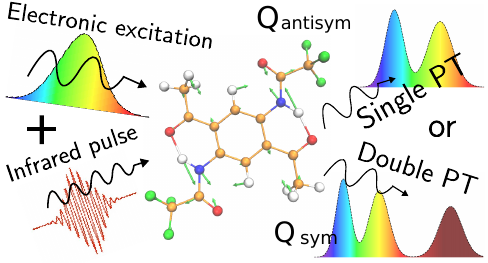}

\end{tocentry}

\begin{abstract}
\noindent 
Single-benzene fluorophores (SBFs) are exceptionally light-weight quantum emitters which have recently been shown to display Excited State Intramolecular Proton Transfer (ESIPT) and highly tunable dual-fluorescence that could be exploited for multiple applications. Here, we employ excited-state molecular dynamics to investigate the ultrafast dynamics of ESIPT in two different SBFs, one of which with a new structure proposed in this work. We find that both single and double ultrafast proton transfers can be triggered by exciting specific molecular vibrations. Strikingly, the kinetic activation of these vibrations -- selected by atomistic, symmetry-based analysis -- can considerably enhance the yield of proton transfer(s), as well as direct the reaction towards either single or double-transfer products. In-depth analysis elucidates the origins of this vibronic control, providing atomistic guidelines through which both the rate and fate of ESIPT could be steered using ultrafast IR-Vis spectroscopies.
\end{abstract}

\begin{figure}[bht!]
  \hspace*{-1.0cm} \includegraphics[scale=0.27]{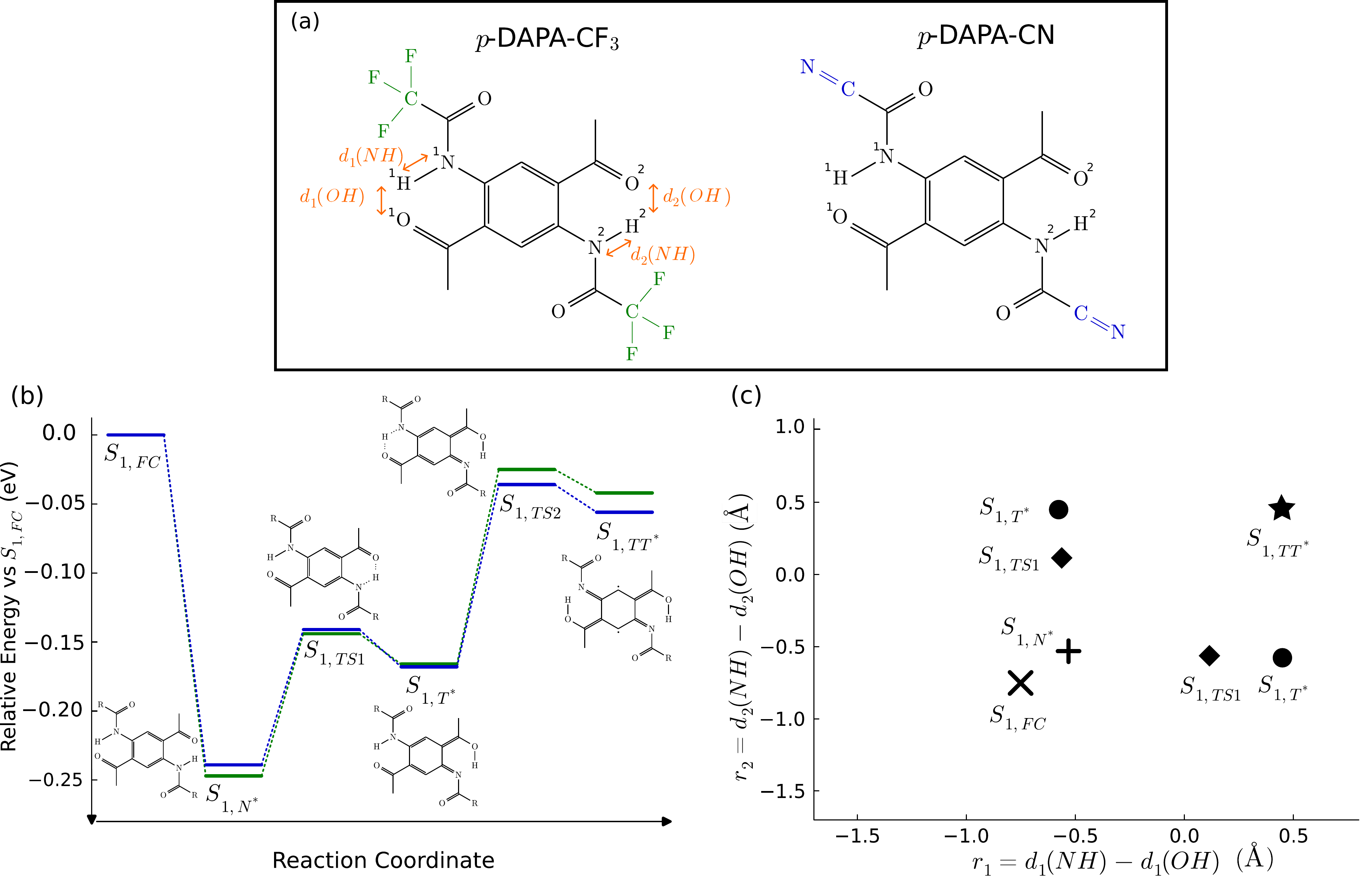}
  \caption{(a) Structure of the studied systems $p$-DAPA-CF$_3$ (left) and the new system proposed in this work $p$-DAPA-CN (right) with the definition of the different proton distances. (b) Schematic Potential Energy Surface for \textit{p}-DAPA-CF$_3$ (green) and \textit{p}-DAPA-CN (blue). The high electron-withdrawing effect of the (CN) substituent reduces the barrier to reach S$_{1,T^*}$. (c) Position in the reaction coordinates plane $(r_1,r_2)$ of the relevant molecular geometries: local minima, Franck-Condon and transition states. }
     \label{fig:CNstruct_PES}
        \end{figure}

Single-benzene fluorophores (SBFs) are a new and promising family of small molecular emitters that are noticeable for their large Stokes shift ($> 12400~$cm$^{-1}$), their exceptionally light molecular weight \cite{kim_relief_2021,chatterjee_meta_2023,choi_urea_fused_2024,huang_precision_2025}, and their (chemically) tunable emission across the full visible spectrum \cite{kim_relief_2021,kim_positional_2024,heo_elucidating_2025}. One explanation for this tuneability in fluorescence is assigned to anti-aromaticity based on the Baird's rule \cite{baird_quantum_1972,karas_bairds_2022}, where the benzene ring becomes stabilized by bond length redistribution in the excited state \cite{kim_relief_2021,yan_photochemistry_2023, xing_proton_2024,chen_meta_2025}. But, besides this intriguing property, the prototypical SBF \textit{para}-diacetylphenylenediamine (\textit{p}-DAPA -- see Fig. \ref{fig:CNstruct_PES}a) -- which is currently recognized as the lightest known red emitter \cite{kim_relief_2021} -- is also known to undergo Excited-State Intramolecular Proton Transfer (ESIPT) \cite{kim_singlebenzene_2023}. Indeed, strongly red-shifted emission from the proton-transferred product \emph{co-exists} with the emission of the non-transferred reactant, making this  -- and other -- SBFs so-called dual-fluorescence emitters \cite{sun_engineering_2024,sun_stimuliresponsive_2025,tian_recent_2025}.

Such dual-emitters are ideal candidates for diverse applications, such as biological cell sensors, medical imaging and the production of organic white-light LEDs for energy efficient lighting. \cite{kim_smallbeautiful_2024,raghava_aminotereph_2024,tian_recent_2025}. However, an additional feature of \textit{p}-DAPA that can be seen in Fig. \ref{fig:CNstruct_PES}a, is that this SBF can in fact undergo  \emph{double-proton transfer}, potentially leading to a novel type of triple-fluorescence, if all three excited state species could be populated. Yet, to date, the doubly proton-transferred state has not be detected, and is thought to be energetically inaccessible under optical excitation from the ground state geometry.    

On a molecular level, the yield of ESIPT in \textit{p}-DAPA has been shown to increase with the addition of electron-withdrawing substituents to its amine groups. However, to date, experimental fluorescence measurements only probe the nanosecond timescale for such systems. At these timescales, the proportion of non-transferred reactant to proton-transferred product follows a thermal probability distribution determined entirely by free energy considerations \cite{kim_singlebenzene_2023}. However in many small molecules, ESIPT is known to be ultrafast ($10-100$ fs), but the ultrafast dynamics immediately following excitation, and their potential contribution to ESIPT and dual fluorescence in SBFs remain unknown. The same is true of the effects that different substituents have on these dynamics. Moreover, one could expect that on the  picosecond timescale, dynamics involve intricate non-equilibrium coupling between the proton motion and the molecular vibrations that this study aims at elucidating.

Motivating this interest in the role of vibrations on ESIPT is not only a longstanding interest in the vibrational regulation of ground-state proton transfer rates \cite{Barbara1992_yq,Hammes_Schiffer1995_rh,Tachikawa1996_nm}, but also the appearance, over the last decade, of a number of experimental techniques that have demonstrated how (external) excitation of molecular vibrations can \emph{controllably} alter reaction rates and outcomes \cite{Delor2014_he,Delor2014_rx,Delor2015_ya}, both in ground and excited states. For example: infrared laser pulses targeting specific IR-active modes via frequency selection have been shown to assist glycine synthesis \cite{Scuderi2020_ao} and can modify charge transport in organic condensed-phase materials \cite{Lynch2011_ub,Bakulin2015_vj}.

Of particular relevance for ESIPT, the combination of UV-Vis pumps and IR pulses has also been shown to re-direct excited state electron transfer in metal-ligand complexes \cite{Lin2009_rz,Yue2015_ny,Delor2017_yo}, and, \textit{inter alia}, the elucidation of the causes of these particular `vibrationally activated' effects has become a subject of considerable theoretical interest \cite{Sola2024, Sola2025, Haoyang2025}. Here, we combine these theoretical and experimental perspectives by investigating the possible vibrational control of ESIPT in SBFs on ultrafast timescales where real-time vibronic dynamics -- and not just thermodynamics -- could play a deciding role \cite{Tomin2010_rd,joshi_excited-state_2021}. Notably, we shall demonstrate that pumping of specific, spectrally resolvable IR modes can cause excited-state dynamics that drive the system \emph{directly} through a concerted double-ESIPT to a \emph{stable} and highly red-shifted emissive product.

\newpage

\textbf{Methodology}. In this work, we focus on the ESIPT dynamics in the prototypical dual-fluorescent system \textit{p}-DAPA-CF$_3$ (Fig.~\ref{fig:CNstruct_PES}~(a)) as well as in a new molecule, that we propose in this work: \textit{p}-DAPA-CN in which the (CF$_3$) moieties are replaced by nitriles (CN) (Fig.~\ref{fig:CNstruct_PES}~(a)). Though this last system has yet to be synthetized, it could, in principle, be obtained using precursors such as phosphoryltrichloride (POCl3) \cite{Zhan2015_mx,Yang2015_ys} and other, as discussed by Miele \textit{et al} \cite{Miele2024_he}.
Based on the experimental work of Kim \textit{et al}. and their static time-dependent density functional theory (TD-DFT) predictions \cite{kim_singlebenzene_2023}, we first determined the minima and saddle point geometries for the proton transfer reactions
in the first singlet excited state using the TD-DFT formalism. We then run molecular dynamics in the first Born-Oppenheimer excited state, S$_1$,
using our own software to integrate Newton's equation of motion coupled with electronic structure software to
get energies and forces (see SI for more details)~\cite{Angiolari2023_pi}.

Before detailing these dynamics, we briefly describe the nearly identical
S$_1$ potential energy surfaces (PESs) in the two systems under study, which have very similar positions for the energy minima and the transition states. The S$_1$ surface shown in Fig.~\ref{fig:CNstruct_PES}~(b) was obtained using TD-DFT with the B3LYP functional \cite{Becke1993_ta,Lee1988_cz} and dispersion correction via D3 method \cite{Grimme2010_tx} using the  6-31++G$^{**}$
basis set as implemented in the Gaussian16 software~\cite{g16}. 
The relevant geometries of the S$_1$ excited surface were obtained by relaxation, starting from geometries close to the different minima and transition states. Simulations are performed for isolated molecules in vacuum, however, we note that the obtained transition energies are similar to those previously reported in the literature with an implicit CHCl$_3$ solvent \cite{kim_singlebenzene_2023}. Other basis sets and functionals have been also been employed, and the details of TD-DFT simulations, and the justifications for our choice of parameters, can be found in Figure~S1 and Tables~(S1-S5) in the Supporting Information (see Figure~S2 for a side note between static calculations and the fluorescence spectrum).

To perform ESIPT dynamics, the system is first promoted vertically from the ground-state minimum to the first excited state S$_1$, which corresponds to the Franck-Condon geometry denoted S$_{1,FC}$ (Fig.~\ref{fig:CNstruct_PES}~(b)). 
The Franck-Condon geometry is not a minimum of the S$_1$ surface and the system will therefore relax towards the first minimum S$_{1,N^*}$ (hence denoted as the N$^*$ state, where the proton has \textit{not} been transferred,  following the notation originally used by Kim et al. \cite{kim_singlebenzene_2023}). 
This N$^*$ state is typically identified as the equilibrium reactant state for the further ESIPT reactions.
To reach the proton's \textit{transferred} state geometry S$_{1,T^*}$ (hence denoted as the T$^*$ state), the system has to overcome a barrier characterized by the transition state with the geometry S$_{1,TS1}$. A second proton transfer yields the S$_{1,TT^*}$ geometry (hence denoted as the TT$^*$ state) via the transition state geometry S$_{1,TS2}$. 

In previous studies on these systems, Kim et al.~\cite{kim_relief_2021, kim_singlebenzene_2023} have correlated the electron-withdrawing ability of the (CF$_3$) moieties with the observed ease of ESIPT. 
Based on this observation, we will also study a hypothetical $p$-DAPA-CN molecule, where the cyanide groups are strongly electron-withdrawing but much lighter and less sterically cumbersome than CF$_3$ groups. Our TD-DFT calculations thereby confirm the importance of the electron-withdrawing effect since the new molecule exhibits a lower energy barrier between S$_{1,N^*}$ and S$_{1,TS1}$ (see Fig.~\ref{fig:CNstruct_PES} and Table~S4). This difference in energy barriers has a significant impact on the ESIPT dynamics, as we demonstrate below.

To analyse the proton displacements, the global reaction coordinates $ r_1 = d_1\text{(NH)} - d_1\text{(OH)}$ and  $ r_2 = d_2\text{(NH)} - d_2\text{(OH)}$ are defined with $d\text{(NH)}$ the distance between the hydrogen and the nitrogen (proton-donor) and $d\text{(OH)}$ the distance between the hydrogen and the oxygen (proton-acceptor) (see Fig.~\ref{fig:CNstruct_PES}~(a)). The positions in the $(r_1, r_2)$-plane of the local energy minima and of the transition states are reported in Fig.~\ref{fig:CNstruct_PES}~(c). 

\textbf{Ultrafast dynamics at different temperatures.} We carried out reaction dynamics simulations in the first excited state using TD-DFT gradients, allowing us to access the first picoseconds of the ESIPT on the S$_1$ PES. We note that the Franck-Condon geometry is significantly higher in energy than the first transition state S$_{1,TS1}$ and even slightly higher than the second one S$_{1,TS2}$ (Fig.~\ref{fig:CNstruct_PES}~(b)). Based on purely energetic considerations, this could allow for ultrafast proton transfer to occur, and potentially even double proton transfer, if the excess Franck-Condon energy is readily available for the protons to go over the corresponding barriers. However, the double transfer process is not experimentally observed \cite{kim_singlebenzene_2023}, which suggests that the energy might be distributed among modes that do not directly contribute to ESIPT, and is lost as heat, before a double transfer can take place.

\begin{figure}[bht!]
  \centering  \includegraphics[scale=0.35]{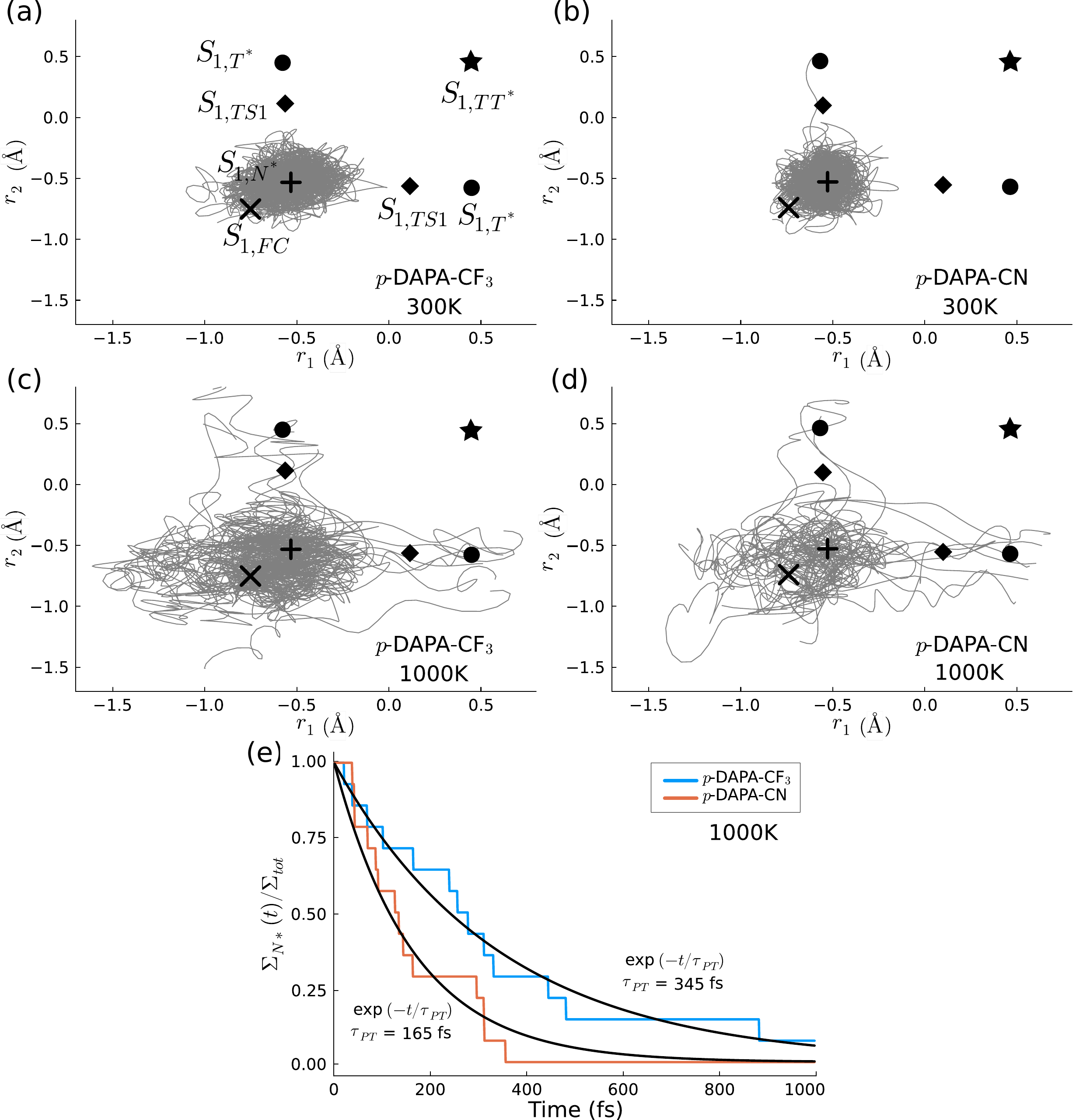}
  \caption{Superimposed plots of 10 trajectories obtained from molecular dynamics projected on the $r_1$-$r_2$ variable plane for (a) $p$-DAPA-CF$_3$ at 300~K, (b) $p$-DAPA-CN at 300~K, (c) $p$-DAPA-CF$_3$ at 1000~K, (d) $p$-DAPA-CN at 1000~K. The stable geometries and transition states are also reported as bullet points of different shapes, (e) proportion of the non-transferred state at 1000~K over time $\Sigma_{N^*}(t)/\Sigma_\text{tot}$.} 
  \label{fig:trajr1r2_tot}
        \end{figure}

We first performed an ensemble of 1~ps-long simulations using as initial conditions the 
positions and velocities obtained from a thermal sampling on the ground state PES at 300 or 1000~K. Details on the ground state sampling procedure can be found in the Supporting Information. To analyse the proton displacements, the reaction coordinates $ r_1$ and $r_2$ are monitored and the trajectories are projected onto these two coordinates for both systems at 300 and 1000~K, as reported in Fig.~\ref{fig:trajr1r2_tot}. \\

At 300~K, only one trajectory over a total of ten for \textit{p}-DAPA-CN shows a single proton transfer and none for \textit{p}-DAPA-CF$_3$ (Fig.~\ref{fig:trajr1r2_tot}~(a-b)). Increasing the temperature (1000~K) enables reactivity, which can be quantified by computing the ratio $\Sigma_{N^*}(t)/\Sigma_\text{tot}$, with $\Sigma_\text{tot}$ the total number of trajectories for a given system and $\Sigma_{N^*}(t)$  the number of trajectories still remaining in the N$^*$ state at the simulation time $t$ (Fig.~\ref{fig:trajr1r2_tot}~(e)). An exponential fit of this quantity yields the proton transfer time at 1000 K $\stau_\text{PT} =345~$fs for $p$-DAPA-CF$_3$ and $\stau_\text{PT} =165~$fs for $p$-DAPA-CN. The associated mean transfer times as well as the sampling uncertainties ($280\pm 145~$fs for $p$-DAPA-CF$_3$ and $159\pm 66~$fs for $p$-DAPA-CN) are calculated in the Supporting Information (see Tables~S6 and S7) . The high temperatures required to observe a proton transfer within the first picosecond after the excitation apparently rule out an ultrafast contribution to the ESIPT mechanism at room temperature, at least within the Born-Oppenheimer approximation. The transfer timescales obtained from the reactive dynamics are also consistent with that estimated from transition state theory for a thermally activated reaction starting in S$_{1,N^*}$ (see Supporting Information). This further reinforces the picture of an initial relaxation to S$_{1,N^*}$ followed by thermally activated ESIPT, rather than an ultrafast transfer process. This also highlights the importance to take into account the multi-dimensional energy distribution, as a simplified one-dimensional reaction coordinate picture could lead to wrongly predict high-yield ultrafast transfer, especially considering that only 40~fs are needed at 0~K for the system to travel from S$_{1,FC}$ to S$_{1,N^*}$. To complete this analysis, we monitored the carbon distances of the benzene ring, but no strong evidence on the role of the benzene in the reactive dynamics can be revealed (Figure~S3 in Supporting Information). The adiabatic energetic gaps were also measured and their variations along typical trajectories can be found in Figure~S4-S7, showing that the ground state and S$_1$ states remain far apart in energy, as well as for the S$_2$ and S$_1$ states maintaining a smaller but significant energy gap, thus justifying the Born-Oppenheimer approximation. 

\newpage

\textbf{Vibrational control.} 
The above results show that the excited state proton transfer is not an ultrafast process at room temperature. We now investigate the possibility to enhance the ESIPT dynamics and possibly the double-transfer process via  vibrational control in the excited state.

To this end, we carried out excited state simulations with different initial conditions. Considering the non-transferred state N$^*$ as the reactant, we performed a normal mode analysis and selected some specific modes with a significant component along the proton transfer coordinate, therefore being likely to contribute to the overcoming of the barrier towards the transferred state T$^*$. The dynamics are then initialized in the N$^*$  equilibrium reactant geometry, but with a velocity boost on the selected mode, giving it a finite amount of kinetic energy.

\begin{figure}[bht!]
    \hspace*{-1.0cm}  \includegraphics[scale=0.26]{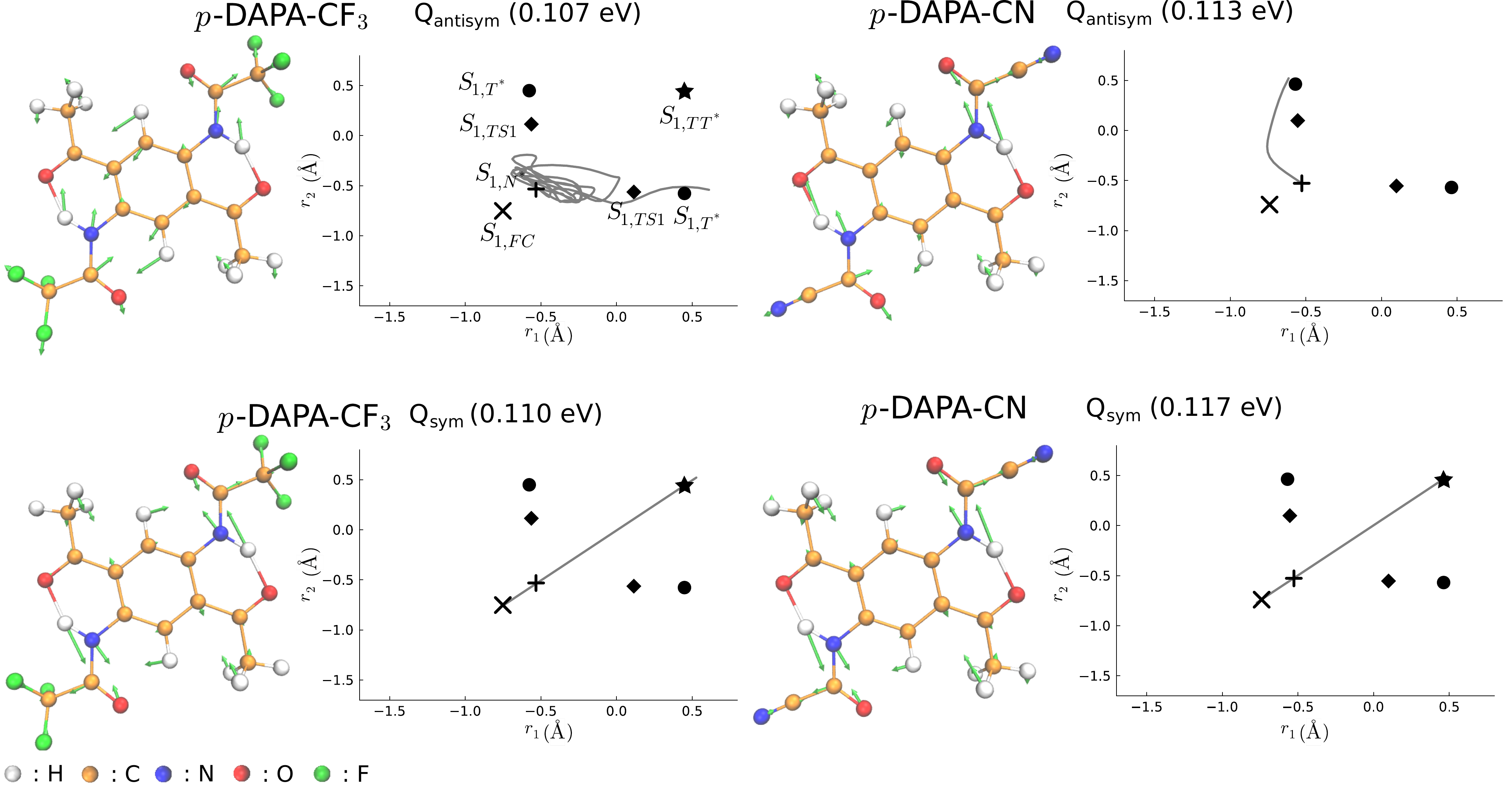}
    \caption{Symmetric and antisymmetric normal mode displacement $Q$ for $p$-DAPA-CF$_3$ and $p$-DAPA-CN with their respective frequencies ($\hbar \omega$) in parenthesis. The symmetric mode leads to a double proton transfer whereas the antisymmetric mode leads to a single proton transfer. Trajectories in the ($r_1$-$r_2$)-plan are represented for the 0~K initial configuration case with the highest vibrational energy in the targeted mode.}
    \label{fig:NMsymantisym}
\end{figure}

For the other normal modes we considered different initial conditions: first, we assigned them zero initial velocity, corresponding to a 0~K (classical) situation with absorption of an infrared radiation resonant with the targeted vibration. 
By carrying out  molecular dynamics with  different single vibration activations (see Supporting Information), two normal modes are identified in both systems as requiring the less energy to promote the proton transfer (Fig.~\ref{fig:NMsymantisym}). These identified vibrations correspond to a symmetric and an antisymmetric mode, that cause significant change in the proton-donor proton-acceptor distances. In the symmetric mode, both protons move simultaneously toward the proton acceptor while in the antisymmetric mode, they move in opposite directions, as illustrated in Fig.~\ref{fig:NMsymantisym}. The selected modes correspond to very similar displacements in both systems, with the most noticeable difference being in the movement of the proton in the antisymmetric mode: whereas in $p$-DAPA-CN, the proton displacement is almost perfectly aligned along the donor-acceptor direction, in $p$-DAPA-CF$_3$, in contrast, the proton motion deviates slightly from this axis and shows a smaller amplitude. This difference is of large impact for vibrational control. \\

Remarkably, when vibrationally targeted, the symmetric mode yields the double proton transfer product TT$^*$ (Fig.~\ref{fig:NMsymantisym}), whereas this product has not been experimentally probed in fluorescence measurements. On the other hand, when activating the antisymmetric mode, only a single proton transfer occurs since, once it has been transferred, the system rearrangement prohibits a second delayed proton transfer to take place. The results of the
simulations with the all but one vibration at rest (0~K) are summarized in Table~\ref{table:vibcontrol_0K}. While a high amount of energy is required to promote ultrafast proton transfer, these runs demonstrate the possibility to activate proton transfer(s) via vibrational control. 

 \begin{table}[bht!]
 \centering
\caption{\textmd{Proton transfer time, $\stau_\text{PT}$ (in fs) under vibrational control at 0~K. \textcolor{red}{X} indicates the absence of proton transfer whereas \textcolor{teal}{$\times 2$} denotes a double proton transfer.}}
\label{table:vibcontrol_0K}
\setlength{\aboverulesep}{0pt}
\setlength{\belowrulesep}{0pt}
\resizebox{0.78\textwidth}{!}{%
\begin{tabular}{c | c | c | c | c}

E$_\text{vib}$ (eV) & {\makecell{$p$-DAPA-CF$_3$\\sym}} & {\makecell{$p$-DAPA-CF$_3$\\antisym}} & {\makecell{$p$-DAPA-CN\\sym}}  & {\makecell{$p$-DAPA-CN\\antisym}} \\
\midrule 2.48 &  34 \textcolor{teal}{($\times 2$)} & 214 & 32 \textcolor{teal}{($\times 2$)} & 13   \\
\midrule
1.22  &  71 \textcolor{teal}{($\times 2$)} & \textcolor{red}{X}  & 66 \textcolor{teal}{($\times 2$)} & 46  \\
\midrule
0.893 &  \textcolor{red}{X} & \textcolor{red}{X}  & \textcolor{red}{X} & 86   \\
\midrule
0.620 &  \textcolor{red}{X} & \textcolor{red}{X}  & \textcolor{red}{X} & \textcolor{red}{X} \\
\end{tabular}%
}

\end{table}

 \begin{table}[ht]
 \centering
\caption{\textmd{Number of transfer under vibrational activation over four different 300~K sampling conditions (in parenthesis) and proton transfer averaged time (in fs)  for the symmetric (sym) and antisymmetric (antisym) modes for $p$-DAPA-CF$_3$ and $p$-DAPA-CN. \textcolor{red}{X} indicates the absence of proton transfer and \textcolor{teal}{$\times 2$} denotes a double proton transfer}}
\label{table:vibcontrol_300K}
\setlength{\aboverulesep}{0pt}
\setlength{\belowrulesep}{0pt}
\resizebox{0.9\textwidth}{!}{%
\begin{tabular}{c |*{2}{c}| *{2}{c}| *{2}{c} |*{2}{c}}
E$_\text{vib}$ (eV) & \multicolumn{2}{c|}{\makecell{$p$-DAPA-CF$_3$\\sym}} &  \multicolumn{2}{c|}{\makecell{$p$-DAPA-CF$_3$\\antisym}}  & \multicolumn{2}{c|}{\makecell{$p$-DAPA-CN\\sym}}  &  \multicolumn{2}{c}{\makecell{$p$-DAPA-CN\\antisym}} \\ 
\midrule 1.22  &  (4/4) & 61 & (3/4) & 71 & (4/4 -- 1\textcolor{teal}{($\times 2$)}) & 62 & (4/4) & 31  \\
\midrule 0.620  & (3/4) & 199 & (3/4) & 342 & (3/4) & 109 & (4/4) & 51  \\
\midrule 0.400  & (2/4) & 73 & (0/4) &  \textcolor{red}{X} & (2/4) & 399 & (4/4) & 82   \\
\midrule 0.200  & (1/4) & 108 & (0/4) &  \textcolor{red}{X} & (1/4) & 141 & (3/4) & 82   \\
\midrule $\hbar \omega$ & (1/4) & 110 & (0/4) &  \textcolor{red}{X} & (1/4) & 143 & (3/4) & 87  \\
\midrule 0  & (0/4) &  \textcolor{red}{X} & (0/4) &  \textcolor{red}{X} & (0/4) &  \textcolor{red}{X} & (0/4) &  \textcolor{red}{X} \\
\end{tabular}%
}

\end{table}

In a second step, we run trajectories with the same vibrational control procedure for the initialization but considering a 300~K thermal equilibrium configuration. In this case, all modes are thermally sampled around the minimum of the excited state, while the identified vibration is additionally targeted by an infrared pulse. As highlighted by the mean proton transfer time averaged over four different thermal samplings (Table~\ref{table:vibcontrol_300K}, see Table~S8 for details), heating up the system largely suppresses the double transfer events, as both the symmetric and the antisymmetric modes now mostly lead to the single transfer product (apart from a single observation in $p$-DAPA-CN) (see Fig.~S9-S10).
Furthermore, thermal fluctuations also lead to a decrease in the amount of energy required to reach the T$^*$ product. Indeed, even a single photon ($\hbar \omega$) absorption can promote the ultrafast transfer that was almost impossible in the first picosecond at such temperatures in the absence of vibrational control, projecting this result to an experimental reality. Remarkably, the proportion of reacting trajectories is significantly higher for $p$-DAPA-CN than for $p$-DAPA-CF$_3$ (for the activation of the antisymmetric mode). This effect can be traced back to two factors: first, the proton transfer barrier is slightly lower in $p$-DAPA-CN, and second, the proton motion in the antisymmetric mode is wider and more closely aligned to the donor-acceptor axis. This observation highlights the subtle balance needed to be able to control the ESIPT process via targeted vibrational infrared excitation .
Finally, we note that the nuclear motion in the performed excited state dynamics is entirely classical, our results therefore do not account for the influence of quantum effects, in particular vibrational zero-point energy and proton tunneling, that might further enhance the reactivity. \\

In conclusion, an excited-state Born-Oppenheimer molecular dynamics study for single-benzene fluorophores has been carried out in order to investigate the ultrafast regime of ESIPT in these systems. 
Dynamics revealed the negligible contribution of the ultrafast dynamics in the ESIPT outcome at room temperature, which leads us to conclude that the process is essentially thermally activated. However, by performing a vibrational activation in the excited state, a symmetric and antisymmetric pair of modes were identified that were able to trigger proton transfer in both $p$-DAPA-CF$_3$ and $p$-DAPA-CN. These normal modes, when vibrationally activated, drive either a double proton transfer (symmetric mode) or a single proton transfer (antisymmetric mode). Thermal fluctuations decrease the amount of externally pumped energy needed to observe vibrationally targeted ESIPT, but the consequent loss of symmetry (on average) inhibits the double proton transfer, indicating the fragility/narrowness of this pathway. A possible way to circumvent this would be to try to engineer more rigidity in future SBFs, though without compromising their appeal as light-weight emitters.      

These findings pave a way for leveraging vibrational activation and temperature for controlling single and double ESIPT for future applications, as well as illustrate the promise of single-benzene fluorophores for the deeper theoretical understanding of multi-ESIPT in general. We suggest that more sophisticated calculations involving quantum nuclear effects would be particularly insightful -- and tractable -- in this emerging class of small molecular fluorophores.\\

\newpage

\begin{acknowledgement}
We thank iSiM (Initiative Sciences et ing\'enierie mol\'eculaires) 
 from the Alliance Sorbonne Universit\'e for funding. 
\end{acknowledgement}

\begin{suppinfo}

TD-DFT parameters, benchmark and results. Details of potential energy surface values. Transition state theory calculation.  Side note on the study of emission spectrum. Dynamics parameters and ground state sampling procedure. Trajectory details. Benzene C-C distances during the dynamics.  Adiabatic energetic gaps between S$_0$ and S$_1$ and between S$_1$ and S$_2$. Detailed transfer time for vibrational control at 300~K. Analysis of the vibrational energy distribution.

\end{suppinfo}

\bibliography{SBF}

@INCOLLECTION{Tomin2010_rd,
  title     = "Proton transfer reactions in the excited electronic state",
  booktitle = "Hydrogen Bonding and Transfer in the Excited State",
  author    = "Tomin, Vladimir I.",
  publisher = "John Wiley \& Sons, Ltd",
 DOI = {10.1002/9780470669143.ch22},
  pages     = "463--523",
  month     =  nov,
  year      =  2010,
  address   = "Chichester, UK"
}

@article{joshi_excited-state_2021,
	title = {Excited-{State} {Intramolecular} {Proton} {Transfer}: {A} {Short} {Introductory} {Review}},
	volume = {26},
	issn = {1420-3049},
	shorttitle = {Excited-{State} {Intramolecular} {Proton} {Transfer}},
	url = {https://www.mdpi.com/1420-3049/26/5/1475},
	doi = {10.3390/molecules26051475},
	language = {en},
	number = {5},
	urldate = {2022-10-11},
	journal = {Molecules},
	author = {Joshi, Hem C. and Antonov, Liudmil},
	month = mar,
	year = {2021},
	pages = {1475},
}

@article{kim_singlebenzene_2023,
	title = {Single‐{Benzene} {Dual}‐{Emitters} {Harness} {Excited}‐{State} {Antiaromaticity} for {White} {Light} {Generation} and {Fluorescence} {Imaging}},
	volume = {62},
	issn = {1433-7851, 1521-3773},
	url = {https://onlinelibrary.wiley.com/doi/10.1002/anie.202302107},
	doi = {10.1002/anie.202302107},
	language = {en},
	number = {20},
	urldate = {2024-11-25},
	journal = {Angewandte Chemie International Edition},
	author = {Kim, Younghun and Kim, Heechan and Son, Jung Bae and Filatov, Michael and Choi, Cheol Ho and Lee, Nam Ki and Lee, Dongwhan},
	month = may,
	year = {2023},
	pages = {e202302107}
}

@article{kim_smallbeautiful_2024,
	title = {Small is {Beautiful}: {Electronic} {Origin} and {Synthetic} {Evolution} of {Single}-{Benzene} {Fluorophores}},
	volume = {57},
	copyright = {https://doi.org/10.15223/policy-029},
	issn = {0001-4842, 1520-4898},
	shorttitle = {Small is {Beautiful}},
	url = {https://pubs.acs.org/doi/10.1021/acs.accounts.3c00605},
	doi = {10.1021/acs.accounts.3c00605},
	language = {en},
	number = {1},
	urldate = {2024-11-25},
	journal = {Accounts of Chemical Research},
	author = {Kim, Heechan and Kim, Younghun and Lee, Dongwhan},
	month = jan,
	year = {2024},
	pages = {140--152}
}

@article{kim_relief_2021,
	title = {Relief of excited-state antiaromaticity enables the smallest red emitter},
	volume = {12},
	issn = {2041-1723},
	url = {https://www.nature.com/articles/s41467-021-25677-2},
	doi = {10.1038/s41467-021-25677-2},
	language = {en},
	number = {1},
	urldate = {2024-11-26},
	journal = {Nature Communications},
	author = {Kim, Heechan and Park, Woojin and Kim, Younghun and Filatov, Michael and Choi, Cheol Ho and Lee, Dongwhan},
	month = sep,
	year = {2021},
	pages = {5409}
}

@article{baird_quantum_1972,
	title = {Quantum organic photochemistry. {II}. {Resonance} and aromaticity in the lowest 3.pi..pi.* state of cyclic hydrocarbons},
	volume = {94},
	issn = {0002-7863, 1520-5126},
	url = {https://pubs.acs.org/doi/abs/10.1021/ja00769a025},
	doi = {10.1021/ja00769a025},
	language = {en},
	number = {14},
	urldate = {2024-11-27},
	journal = {Journal of the American Chemical Society},
	author = {Baird, N. Colin},
	month = jul,
	year = {1972},
	pages = {4941--4948}
}

@article{huang_precision_2025,
	title = {Precision {Molecular} {Engineering} of {Compact} {Near}-{Infrared} {Fluorophores}},
	copyright = {https://doi.org/10.15223/policy-029},
	issn = {0002-7863, 1520-5126},
	url = {https://pubs.acs.org/doi/10.1021/jacs.4c16087},
	doi = {10.1021/jacs.4c16087},
	language = {en},
	urldate = {2025-02-07},
	journal = {Journal of the American Chemical Society},
	author = {Huang, Rongrong and Qiao, Qinglong and Seah, Deborah and Shen, Tianruo and Wu, Xia and De Moliner, Fabio and Wang, Chao and Ding, Nannan and Chi, Weijie and Sun, Huaming and Vendrell, Marc and Xu, Zhaochao and Fang, Yu and Liu, Xiaogang},
	month = feb,
	year = {2025},
	pages = {jacs.4c16087}
}

@misc{chen_meta_2025,
	title = {Meta effect enables redder and larger {Stokes} shift chromophores by enhanced aromaticity reversal},
	copyright = {https://creativecommons.org/licenses/by-nc-nd/4.0/},
	url = {https://chemrxiv.org/engage/chemrxiv/article-details/67bda1fffa469535b9dad3c4},
	doi = {10.26434/chemrxiv-2025-56x1p},
	language = {en},
	urldate = {2025-02-28},
	author = {Chen, Cheng and Liu, Jiawei and Myasnyanko, Ivan N. and Rudik, Daniil I. and Adams, Anita H. and Walker, Alice R. and Baranov, Mikhail S. and Fang, Chong},
	month = feb,
	year = {2025}
}

@article{chatterjee_meta_2023,
	title = {\textit{meta} -{Fluorophores}: an uncharted ocean of opportunities},
	volume = {59},
	issn = {1359-7345, 1364-548X},
	shorttitle = {\textit{meta} -{Fluorophores}},
	url = {https://xlink.rsc.org/?DOI=D3CC04182D},
	doi = {10.1039/D3CC04182D},
	language = {en},
	number = {97},
	urldate = {2024-11-26},
	journal = {Chemical Communications},
	author = {Chatterjee, Tanmay and Mandal, Mrinal and Mardanya, Sukumar and Singh, Manjeev and Saha, Arijit and Ghosh, Swarnali and Mandal, Prasun K.},
	year = {2023},
	pages = {14370--14386}
}

@article{sun_engineering_2024,
	title = {Engineering {Tunable} {Ratiometric} {Dual} {Emission} in {Single} {Emitter}‐based {Amorphous} {Systems}},
	volume = {63},
	issn = {1433-7851, 1521-3773},
	url = {https://onlinelibrary.wiley.com/doi/10.1002/anie.202318159},
	doi = {10.1002/anie.202318159},
	language = {en},
	number = {10},
	urldate = {2025-06-24},
	journal = {Angewandte Chemie International Edition},
	author = {Sun, Hao and He, Menglu and Baryshnikov, Glib V. and Wu, Bin and Valiev, Rashid R. and Shen, Shen and Zhang, Man and Xu, Xiaoyan and Li, Zhongyu and Liu, Guofeng and Ågren, Hans and Zhu, Liangliang},
	month = mar,
	year = {2024},
	pages = {e202318159}
}

@article{karas_bairds_2022,
	title = {Baird’s rules at the tipping point},
	volume = {14},
	issn = {1755-4330, 1755-4349},
	url = {https://www.nature.com/articles/s41557-022-00988-z},
	doi = {10.1038/s41557-022-00988-z},
	language = {en},
	number = {7},
	urldate = {2024-11-27},
	journal = {Nature Chemistry},
	author = {Karas, Lucas J. and Wu, Judy I.},
	month = jul,
	year = {2022},
	pages = {723--725},
    DOI = {10.1038/s41557-022-00988-z},
}

@article{choi_urea_fused_2024,
	title = {Urea-fused and π-extended single-benzene fluorophores with ultralarge {Stokes} shifts},
	volume = {60},
	copyright = {http://rsc.li/journals-terms-of-use},
	issn = {1359-7345, 1364-548X},
	url = {https://xlink.rsc.org/?DOI=D4CC03389B},
	doi = {10.1039/d4cc03389b},
	language = {en},
	number = {68},
	urldate = {2025-07-13},
	journal = {Chemical Communications},
	author = {Choi, Taehyeon and Kim, Heechan and Kim, Younghun and Lee, Dongwhan},
	year = {2024},
	note = {Publisher: Royal Society of Chemistry (RSC)},
	pages = {9105--9108}
}

@article{heo_elucidating_2025,
	title = {Elucidating the molecular structural origin of efficient emission across solid and solution phases of single benzene fluorophores},
	volume = {16},
	copyright = {https://creativecommons.org/licenses/by-nc-nd/4.0},
	issn = {2041-1723},
	url = {https://www.nature.com/articles/s41467-025-60316-0},
	doi = {10.1038/s41467-025-60316-0},
	language = {en},
	number = {1},
	urldate = {2025-07-14},
	journal = {Nature Communications},
	author = {Heo, Jung-Moo and Park, Jihyun and Flórez-Angarita, Maria F. and Wang, Liangxuan and Yu, Changhoon and Choi, Jinho and Woo, Hochul and Milián-Medina, Begoña and Matzger, Adam J. and Kwon, Min Sang and Gierschner, Johannes and Kim, Jinsang},
	month = jul,
	year = {2025},
	note = {Publisher: Springer Science and Business Media LLC},
}

@article{raghava_aminotereph_2024,
	title = {Amino‐{Terephthalonitrile} and {Amino}‐{Terephthalate}‐{Based} {Single} {Benzene} {Fluorophores} – {Compact} {Color} {Tunable} {Molecular} {Dyes} for {Bioimaging} and {Bioanalysis}},
	volume = {19},
	copyright = {http://onlinelibrary.wiley.com/termsAndConditions\#vor},
	issn = {1861-4728, 1861-471X},
	url = {https://aces.onlinelibrary.wiley.com/doi/10.1002/asia.202400898},
	doi = {10.1002/asia.202400898},
	language = {en},
	number = {23},
	urldate = {2025-07-14},
	journal = {Chemistry – An Asian Journal},
	author = {Raghava, Tanya and Banerjee, Subhadeep},
	month = dec,
	year = {2024},
	note = {Publisher: Wiley}
}

@article{sun_stimuliresponsive_2025,
	title = {Stimuli‐{Responsive} {Dual}‐{Emission} {Property} of {Single}‐{Luminophore}‐{Based} {Materials}},
	volume = {35},
	copyright = {http://onlinelibrary.wiley.com/termsAndConditions\#vor},
	issn = {1616-301X, 1616-3028},
	url = {https://advanced.onlinelibrary.wiley.com/doi/10.1002/adfm.202415400},
	doi = {10.1002/adfm.202415400},
	language = {en},
	number = {7},
	urldate = {2025-07-14},
	journal = {Advanced Functional Materials},
	author = {Sun, Hao and Shen, Shen and Li, Chenzi and Yu, Wanting and Xie, Qishan and Wu, Dayu and Zhu, Liangliang},
	month = feb,
	year = {2025},
	note = {Publisher: Wiley}
}

@article{xing_proton_2024,
	title = {Proton transfer induced excited-state aromaticity gain for chromophores with maximal {Stokes} shifts},
	volume = {15},
	copyright = {http://creativecommons.org/licenses/by/3.0/},
	issn = {2041-6520, 2041-6539},
	url = {https://xlink.rsc.org/?DOI=D4SC04692G},
	doi = {10.1039/d4sc04692g},
	language = {en},
	number = {43},
	urldate = {2025-07-14},
	journal = {Chemical Science},
	author = {Xing, Dong and Glöcklhofer, Florian and Plasser, Felix},
	year = {2024},
	note = {Publisher: Royal Society of Chemistry (RSC)},
	pages = {17918--17926}
}

@article{yan_photochemistry_2023,
	title = {Photochemistry {Driven} by {Excited}‐{State} {Aromaticity} {Gain} or {Antiaromaticity} {Relief}},
	volume = {29},
	copyright = {http://creativecommons.org/licenses/by/4.0/},
	issn = {0947-6539, 1521-3765},
	url = {https://chemistry-europe.onlinelibrary.wiley.com/doi/10.1002/chem.202203748},
	doi = {10.1002/chem.202203748},
	language = {en},
	number = {19},
	urldate = {2025-07-14},
	journal = {Chemistry – A European Journal},
	author = {Yan, Jiajie and Slanina, Tomáš and Bergman, Joakim and Ottosson, Henrik},
	month = apr,
	year = {2023},
	note = {Publisher: Wiley}
}

@article{kim_positional_2024,
	title = {Positional effects of electron-donating and withdrawing groups on the photophysical properties of single benzene fluorophores},
	volume = {60},
	copyright = {http://rsc.li/journals-terms-of-use},
	issn = {1359-7345, 1364-548X},
	url = {https://xlink.rsc.org/?DOI=D4CC04451G},
	doi = {10.1039/d4cc04451g},
	language = {en},
	number = {100},
	urldate = {2025-07-14},
	journal = {Chemical Communications},
	author = {Kim, Dopil and Kim, Jun Yeong and Kim, Haein and Jeong, Eunjin and Lee, Minhyuk and Kim, Dongwook and Kim, JunWoo and Park, Myung Hwan and Kim, Min},
	year = {2024},
	note = {Publisher: Royal Society of Chemistry (RSC)},
	pages = {14956--14959}
}

@article{tian_recent_2025,
	title = {Recent progress in organic dual-emission materials for white organic light emitting diodes ({WOLEDs}) from single small-molecule components},
	volume = {13},
	copyright = {http://rsc.li/journals-terms-of-use},
	issn = {2050-7526, 2050-7534},
	url = {https://xlink.rsc.org/?DOI=D5TC00981B},
	doi = {10.1039/d5tc00981b},
	language = {en},
	number = {22},
	urldate = {2025-07-14},
	journal = {Journal of Materials Chemistry C},
	author = {Tian, Xiangbin and Xia, Yan and Li, Jie and Sun, Jing and Wang, Hua},
	year = {2025},
	note = {Publisher: Royal Society of Chemistry (RSC)},
	pages = {11017--11039}
}

@ARTICLE{Angiolari2023_pi,
  title    = "Environmental and nuclear quantum effects on double proton
              transfer in the guanine-cytosine base pair",
  author   = "Angiolari, Federica and Huppert, Simon and Pietrucci, Fabio and
              Spezia, Riccardo",
  journal  = "The Jounal of Physical Chemistry Letters",
  volume   =  14,
  number   =  22,
  pages    = "5102--5108",
  month    =  jun,
  year     =  2023,
  language = "en"
}

@misc{g16,
author={M. J. Frisch and G. W. Trucks and H. B. Schlegel and G. E. Scuseria and M. A. Robb and J. R. Cheeseman and G. Scalmani and V. Barone and G. A. Petersson and H. Nakatsuji and X. Li and M. Caricato and A. V. Marenich and J. Bloino and B. G. Janesko and R. Gomperts and B. Mennucci and H. P. Hratchian and J. V. Ortiz and A. F. Izmaylov and J. L. Sonnenberg and D. Williams-Young and F. Ding and F. Lipparini and F. Egidi and J. Goings and B. Peng and A. Petrone and T. Henderson and D. Ranasinghe and V. G. Zakrzewski and J. Gao and N. Rega and G. Zheng and W. Liang and M. Hada and M. Ehara and K. Toyota and R. Fukuda and J. Hasegawa and M. Ishida and T. Nakajima and Y. Honda and O. Kitao and H. Nakai and T. Vreven and K. Throssell and Montgomery, {Jr.}, J. A. and J. E. Peralta and F. Ogliaro and M. J. Bearpark and J. J. Heyd and E. N. Brothers and K. N. Kudin and V. N. Staroverov and T. A. Keith and R. Kobayashi and J. Normand and K. Raghavachari and A. P. Rendell and J. C. Burant and S. S. Iyengar and J. Tomasi and M. Cossi and J. M. Millam and M. Klene and C. Adamo and R. Cammi and J. W. Ochterski and R. L. Martin and K. Morokuma and O. Farkas and J. B. Foresman and D. J. Fox},
title={Gaussian˜16 {R}evision {C}.01},
year={2016},
note={Gaussian Inc. Wallingford CT}
}

@ARTICLE{Miele2024_he,
  title     = "Chemoselective synthesis of cyanoformamides from isocyanates and
               a highly reactive nitrile anion \textit{reservoir}",
  author    = "Miele, Margherita and Castoldi, Laura and Roller-Prado,
               Alexander and Pisano, Luisa and Pace, Vittorio",
  journal   = "European J. Org. Chem.",
  publisher = "Wiley",
  volume    =  27,
  number    =  39,
  month     =  oct,
  year      =  2024,
  copyright = "http://creativecommons.org/licenses/by/4.0/",
  language  = "en"
}

@ARTICLE{Yang2015_ys,
  title     = "{POCl3-mediated} reaction of 1-acyl-1-carbamoyl oximes: a new
               entry to cyanoformamides",
  author    = "Yang, Jiming and Xiang, Dexuan and Zhang, Rui and Zhang, Ning
               and Liang, Yongjiu and Dong, Dewen",
  journal   = "Org. Lett.",
  publisher = "American Chemical Society (ACS)",
  volume    =  17,
  number    =  4,
  pages     = "809--811",
  month     =  feb,
  year      =  2015,
  language  = "en"
}

@ARTICLE{Zhan2015_mx,
  title     = "{PhI(OAc)$_{2}$-promoted} metal-free oxidation of
               2-oxoaldehydes: a facile one-pot synthesis of cyanoformamides",
  author    = "Zhan, Zhen and Cheng, Xu and Zheng, Yang and Ma, Xiaojun and
               Wang, Xiaoyu and Hai, Li and Wu, Yong",
  journal   = "RSC Adv.",
  publisher = "Royal Society of Chemistry (RSC)",
  volume    =  5,
  number    =  101,
  pages     = "82800--82803",
  year      =  2015,
  language  = "en"
}

@ARTICLE{Hammes_Schiffer1995_rh,
  title     = "Vibrationally enhanced proton transfer",
  author    = "Hammes-Schiffer, Sharon and Tully, John C",
  journal   = "J. Phys. Chem.",
  publisher = "American Chemical Society (ACS)",
  volume    =  99,
  number    =  16,
  pages     = "5793--5797",
  month     =  apr,
  year      =  1995,
  language  = "en"
}

@ARTICLE{Tachikawa1996_nm,
  title     = "Dynamics of the vibrational mode-specific proton transfer
               reaction {NH3+($\nu$1}) + {NH3} $\rightarrow$ {NH2} + {NH4+}: ab
               initio {MO} and classical trajectory studies",
  author    = "Tachikawa, Hiroto",
  journal   = "Chem. Phys.",
  publisher = "Elsevier BV",
  volume    =  211,
  number    = "1-3",
  pages     = "305--312",
  month     =  nov,
  year      =  1996,
  language  = "en"
}

@ARTICLE{Delor2015_ya,
  title     = "On the mechanism of vibrational control of light-induced charge
               transfer in donor-bridge-acceptor assemblies",
  author    = "Delor, Milan and Keane, Theo and Scattergood, Paul A and
               Sazanovich, Igor V and Greetham, Gregory M and Towrie, Michael
               and Meijer, Anthony J H M and Weinstein, Julia A",
  journal   = "Nat. Chem.",
  publisher = "Springer Science and Business Media LLC",
  volume    =  7,
  number    =  9,
  pages     = "689--695",
  month     =  sep,
  year      =  2015,
  language  = "en"
}

@ARTICLE{Delor2014_he,
  title     = "Toward control of electron transfer in donor-acceptor molecules
               by bond-specific infrared excitation",
  author    = "Delor, Milan and Scattergood, Paul A and Sazanovich, Igor V and
               Parker, Anthony W and Greetham, Gregory M and Meijer, Anthony J
               H M and Towrie, Michael and Weinstein, Julia A",
  journal   = "Science",
  publisher = "American Association for the Advancement of Science (AAAS)",
  volume    =  346,
  number    =  6216,
  pages     = "1492--1495",
  month     =  dec,
  year      =  2014,
doi={10.1126/science.1259995},
  language  = "en"
}

@ARTICLE{Delor2014_rx,
  title     = "Dynamics of ground and excited state vibrational relaxation and
               energy transfer in transition metal carbonyls",
  author    = "Delor, Milan and Sazanovich, Igor V and Towrie, Michael and
               Spall, Steven J and Keane, Theo and Blake, Alexander J and
               Wilson, Claire and Meijer, Anthony J H M and Weinstein, Julia A",
  journal   = "J. Phys. Chem. B",
  publisher = "American Chemical Society (ACS)",
  volume    =  118,
  number    =  40,
  pages     = "11781--11791",
  month     =  oct,
  year      =  2014,
  language  = "en"
}

@ARTICLE{Lin2009_rz,
  title     = "Modulating unimolecular charge transfer by exciting bridge
               vibrations",
  author    = "Lin, Zhiwei and Lawrence, Candace M and Xiao, Dequan and Kireev,
               Victor V and Skourtis, Spiros S and Sessler, Jonathan L and
               Beratan, David N and Rubtsov, Igor V",
  journal   = "J. Am. Chem. Soc.",
  publisher = "American Chemical Society (ACS)",
  volume    =  131,
  number    =  50,
  pages     = "18060--18062",
  month     =  dec,
  year      =  2009,
  language  = "en"
}

@ARTICLE{Yue2015_ny,
  title     = "Electron transfer rate modulation in a compact {Re(I})
               donor-acceptor complex",
  author    = "Yue, Yuankai and Grusenmeyer, Tod and Ma, Zheng and Zhang, Peng
               and Schmehl, Russell H and Beratan, David N and Rubtsov, Igor V",
  journal   = "Dalton Trans.",
  publisher = "Royal Society of Chemistry (RSC)",
  volume    =  44,
  number    =  18,
  pages     = "8609--8616",
  month     =  may,
  year      =  2015,
  language  = "en"
}

@ARTICLE{Barbara1992_yq,
  title     = "Vibrational modes and the dynamic solvent effect in electron and
               proton transfer",
  author    = "Barbara, P F and Walker, G C and Smith, T P",
  journal   = "Science",
  publisher = "American Association for the Advancement of Science (AAAS)",
  volume    =  256,
  number    =  5059,
  pages     = "975--981",
  month     =  may,
  year      =  1992,
  language  = "en"
}

@ARTICLE{Lynch2011_ub,
  title     = "On the role of high-frequency intramolecular vibrations in
               ultrafast back-electron transfer reactions",
  author    = "Lynch, Michael S and Van Kuiken, Benjamin E and Daifuku,
               Stephanie L and Khalil, Munira",
  journal   = "J. Phys. Chem. Lett.",
  publisher = "American Chemical Society (ACS)",
  volume    =  2,
  number    =  17,
  pages     = "2252--2257",
  month     =  sep,
  year      =  2011,
  language  = "en"
}

@ARTICLE{Bakulin2015_vj,
  title     = "Mode-selective vibrational modulation of charge transport in
               organic electronic devices",
  author    = "Bakulin, Artem A and Lovrincic, Robert and Yu, Xi and Selig,
               Oleg and Bakker, Huib J and Rezus, Yves L A and Nayak, Pabitra K
               and Fonari, Alexandr and Coropceanu, Veaceslav and Br{\'e}das,
               Jean-Luc and Cahen, David",
  journal   = "Nat. Commun.",
  publisher = "Springer Science and Business Media LLC",
  volume    =  6,
  number    =  1,
  pages     = "7880",
  month     =  aug,
  year      =  2015,
  copyright = "https://creativecommons.org/licenses/by/4.0",
  language  = "en"
}

@ARTICLE{Delor2017_yo,
  title     = "Directing the path of light-induced electron transfer at a
               molecular fork using vibrational excitation",
  author    = "Delor, Milan and Archer, Stuart A and Keane, Theo and Meijer,
               Anthony J H M and Sazanovich, Igor V and Greetham, Gregory M and
               Towrie, Michael and Weinstein, Julia A",
  journal   = "Nat. Chem.",
  publisher = "Springer Science and Business Media LLC",
  volume    =  9,
  number    =  11,
  pages     = "1099--1104",
  month     =  nov,
  year      =  2017,
  language  = "en"
}

@ARTICLE{Scuderi2020_ao,
  title     = "Infrared-assisted synthesis of prebiotic Glycine",
  author    = "Scuderi, Debora and P{\'e}rez-Mellor, Ariel and Lemaire,
               Jo{\"e}l and Indrajith, Suvasthika and Bardaud, Jean-Xavier and
               Largo, Antonio and Jeanvoine, Yannick and Spezia, Riccardo",
  journal   = "Chemphyschem",
  publisher = "Wiley",
  volume    =  21,
  number    =  6,
  pages     = "503--509",
  month     =  mar,
  year      =  2020,
  copyright = "http://onlinelibrary.wiley.com/termsAndConditions\#vor",
  language  = "en"
}

@article{Haoyang2025,
  title = {Steering Ultrafast Photochemical Reactions Beyond the Transition-State Limit by Tailored Field},
  url = {http://dx.doi.org/10.26434/chemrxiv-2025-zsqqm},
  DOI = {10.26434/chemrxiv-2025-zsqqm},
  publisher = {American Chemical Society (ACS)},
  author = {Haoyang,  Xu and Luxiang,  Zhu and Feng,  Yan and Meifang,  Zhu and Jin,  Wen},
  year = {2025},
  month = Nov 
}

@article{Sola2024,
  title = {Absolute control over the quantum yield of a photodissociation reaction mediated by nonadiabatic couplings},
  volume = {15},
  ISSN = {2041-6539},
  url = {http://dx.doi.org/10.1039/d4sc03235g},
  DOI = {10.1039/d4sc03235g},
  number = {37},
  journal = {Chemical Science},
  publisher = {Royal Society of Chemistry (RSC)},
  author = {Sola,  Ignacio R. and García-Vela,  Alberto},
  year = {2024},
  pages = {15255–15262}
}

@article{Sola2025,
  title = {Preparing Superposition States to Modify the Spectra and to Achieve Complete Selectivity in Photodissociation Reactions},
  volume = {21},
  ISSN = {1549-9626},
  url = {http://dx.doi.org/10.1021/acs.jctc.5c00655},
  DOI = {10.1021/acs.jctc.5c00655},
  number = {15},
  journal = {Journal of Chemical Theory and Computation},
  publisher = {American Chemical Society (ACS)},
  author = {Sola,  Ignacio R. and García-Vela,  Alberto},
  year = {2025},
  month = July,
  pages = {7267–7278}
}

@ARTICLE{Becke1993_ta,
  title     = "Density-functional thermochemistry. {III}. The role of exact
               exchange",
  author    = "Becke, Axel D",
  journal   = "J. Chem. Phys.",
  publisher = "AIP Publishing",
  volume    =  98,
  number    =  7,
  pages     = "5648--5652",
  month     =  apr,
  year      =  1993,
  language  = "en"
}

@ARTICLE{Lee1988_cz,
  title     = "Development of the {Colle-Salvetti} correlation-energy formula
               into a functional of the electron density",
  author    = "Lee, C and Yang, W and Parr, R G",
  journal   = "Phys. Rev. B Condens. Matter",
  publisher = "American Physical Society (APS)",
  volume    =  37,
  number    =  2,
  pages     = "785--789",
  month     =  jan,
  year      =  1988,
  copyright = "http://link.aps.org/licenses/aps-default-license",
  language  = "en"
}

@ARTICLE{Grimme2010_tx,
  title     = "A consistent and accurate ab initio parametrization of density
               functional dispersion correction ({DFT-D}) for the 94 elements
               {H-Pu}",
  author    = "Grimme, Stefan and Antony, Jens and Ehrlich, Stephan and Krieg,
               Helge",
  journal   = "J. Chem. Phys.",
  publisher = "AIP Publishing",
  volume    =  132,
  number    =  15,
  pages     = "154104",
  month     =  apr,
  year      =  2010,
  language  = "en"
}

\end{document}


\section{TD-DFT parameters and results}

Ab initio calculations performed with the Time-Dependent Density Functional Theory (TD-DFT) method are here presented. These calculations are carried out with Gaussian16 \cite{g16}.\\

\begin{figure}[bht!]
    \centering
 \includegraphics[width=0.6\linewidth]{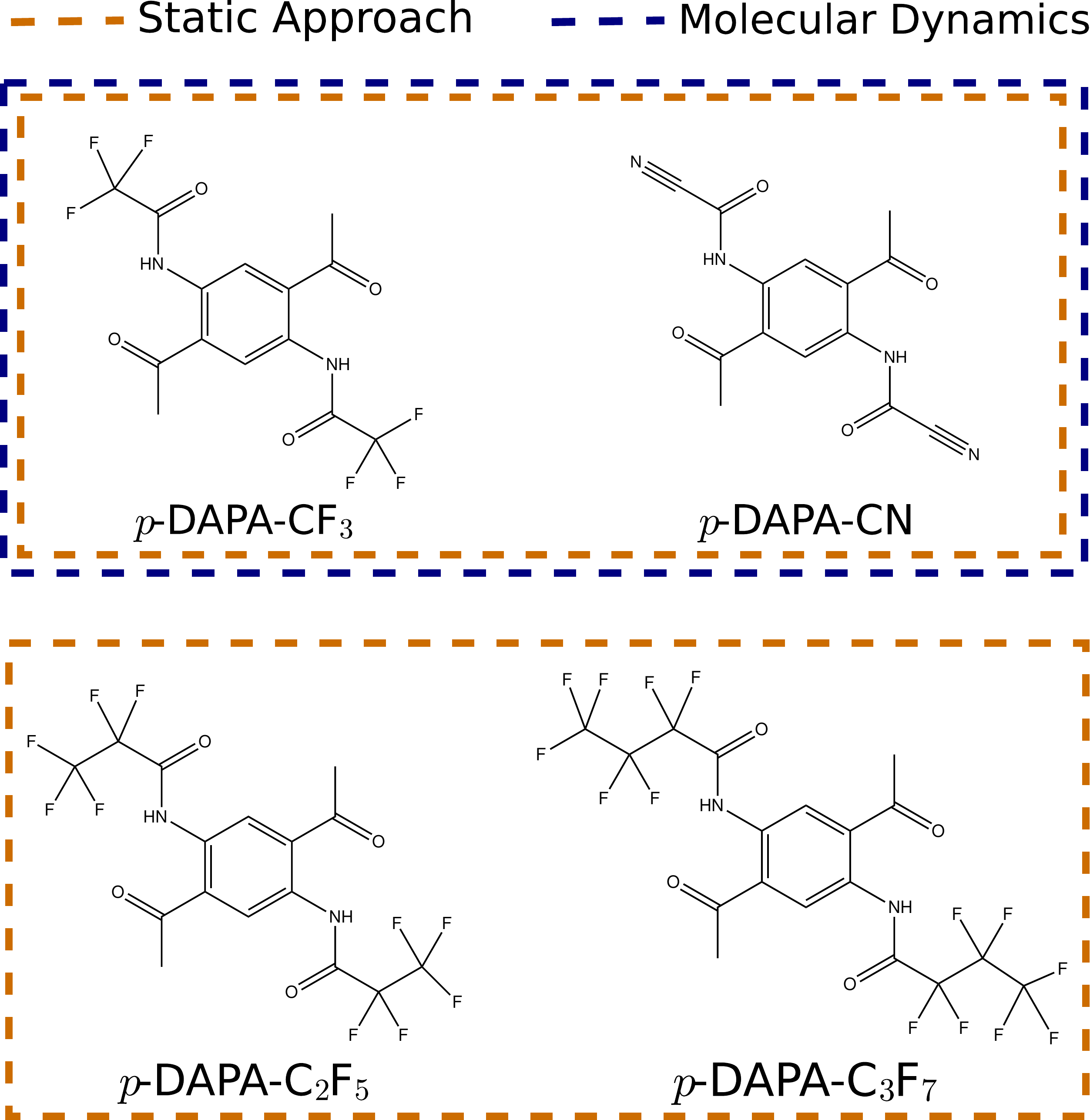}
    \caption{Structure of the studied systems with the static approach. Molecular dynamics also performed in the excited state for \textit{p}-DAPA-CF$_3$ and \textit{p}-DAPA-CN and presented in the main text.}
    \label{fig_chap06:approach_molecules}
\end{figure}

Previous results of the literature for $p$-DAPA-CF$_3$ were obtained at the B3LYP-D3/6-31++G** level including solvent effects \cite{kim_singlebenzene_2023}. In this work, we checked these results at the B3LYP-D3/6-31++G** level without solvent by optimizing reported geometries of the ground state S$_0$ as well as S$_{1,N^*}$, S$_{1,T^*}$, S$_{1,TT^*}$ and transition states S$_{1,TS1}$ and S$_{1,TS2}$. For molecular dynamics, solvent effects include a complex dynamical character that is not the core of the present work, justifying the study of an isolated system. \\

Although these results are consistent with the experimental values, other parameters have been probed that might have enabled computationally less demanding molecular dynamics simulations. As a first step, Density Functional Tight-Binding (DFTB) calculations were carried out to calculate the optimized geometries of the S$_1$ surface. However, we observed that the excited state ranking switched depending on the stable geometry, suggesting crossings between adiabatic states. It thus ruled out DFTB as unreliable for the application to molecular dynamics of the systems under study. \\

\section{Calculation of the S$_1$ Surface}
\label{chap_06:section_static}
\subsection*{Study of other basis sets and functionals}

TD-DFT basis sets and functionals were also scanned for the \textit{p}-DAPA-CF$_3$ system. The transferred state T$^*$ was not found while optimizing the geometry on the excited state for B3LYP-D3/6-31+G ; B3LYP-D3/6-31G ; B3LYPD3/6-21G** ; B3LYP-D3/6-21G ; BLYP-D3/6-21G; B3LYP-D3/3-21+G ; B3LYP-D3/3-21G ; PBE/3-21G ; PBE-D3/3-21G ; PW91/3-21G ; RevTPSS/3-21G ; TPSSh/3-21G, thus also ruling out these computationally lighter choices of parameters. Results are reported in Table~\ref{chap_06:table_dftsmallbasis}. From these values, energetic barriers are calculated in eV and reported in Table~\ref{chap_06:table_dftsmallbasis_barrier}.

   \begin{table}[ht] 
\caption{Scan of other TD-DFT parameters. Energies are in eV and calculated for {\textit{p}-DAPA-CF$_3$}. Energies are relative to the optimized ground state geometry $S_0$.}
\label{chap_06:table_dftsmallbasis}
\begin{minipage}{1.0\textwidth}
\centering
\begin{tabular}{|c|c|c|c|c|c|c|}
\hline
\, & \boldmath$S_{1,FC}$ & \boldmath$S_{1,N^*}$ & \boldmath$S_{1,TS1}$ & \boldmath$S_{1,T^*}$  & \boldmath$S_{1,TS2}$& \boldmath$S_{1,TT}$  \\
\hline \textbf{B3LYP-D3/6-31++G**} & $3.141$ & $2.894$ & $2.997$ & $2.975$ & $3.116$ & $3.099$ \\
\hline \textbf{B3LYP-D3/6-31G(3df, 3pd)} & 3.139&2.869  &2.992  &2.977  & 3.128 &3.113 \\
\hline \textbf{B3LYP/6-31G*} & 3.138 & 2.908 & 3.085 & 3.080 & 3.306 & 3.302 \\
\hline \textbf{B3LYP-D2/6-31G*} & 3.156 & 2.914 &3.098  & 3.090 & 3.322 &3.321  \\
\hline \textbf{B3LYP-D3/6-31G*} & 3.140& 2.908 & 3.093 & 3.080 & 3.313 & 3.303  \\
\hline \textbf{B3LYP-D3(BJ)/6-31G*} & 3.136 & 2.907 & 3.090 & 3.074 & 3.303 & 3.290   \\
\hline \textbf{wB97XD/6-31G*} &3.625 & 3.342 & 3.534 & 3.513 & 3.809 & 3.806   \\
\hline
\end{tabular}

\end{minipage}

\end{table}

  \begin{table}[ht] 
\caption{Scan of other TD-DFT parameters. Barrier energies are in eV and calculated for \textit{p}-DAPA-CF$_3$.}
\label{chap_06:table_dftsmallbasis_barrier}
\begin{minipage}{1.0\textwidth}
\centering
\hspace*{-1.6cm}  \begin{tabular}{|c|c|c|c|c|c|c|}
\hline
\, & $E(S_{1,T^*}) - E(S_{1,N^*})$ & $E(S_{1,TS1}) - E(S_{1,N^*})$ & $E(S_{1,TS1}) - E(S_{1,T^*})$ \\
\hline \textbf{B3LYP-D3/6-31++G**} & $0.0810$ & $0.103$ & $0.0221$\\
\hline \textbf{B3LYP-D3/6-31G(3df, 3pd)} & 0.108 & 0.123  & 0.0148\\
\hline \textbf{B3LYP/6-31G*} & 0.171  & 0.177  & 0.00566 \\
\hline  \textbf{B3LYP-D2/6-31G*} & 0.176  & 0.184 & 0.00801 \\
\hline \textbf{B3LYP-D3/6-31G*} & 0.173  & 0.185 & 0.0123 \\
\hline \textbf{B3LYP-D3(BJ)/6-31G*} & 0.167  & 0.183 & 0.0157 \\
\hline \textbf{wB97XD/6-31G*} & 0.172  & 0.193  & 0.0211 \\
\hline
\end{tabular}

\end{minipage}

\end{table}

We clearly see that numerically cheaper parameters and B3LYP-D3/6-31G(3df, 3pd) significantly change the barrier energies, leading to a different picture concerning the S$_1$ surface and consecutively for the dynamics propagated onto this surface. We therefore chose to keep the B3LYP-D3/6-31++G** parameters.

\subsection*{TD-DFT results}

We first check the energetic landscape of the molecule. Since the seminal experimental paper provides the definition of a reaction coordinate and a PES for the targeted first excited state \cite{kim_singlebenzene_2023}, we first verify that our results are consistent with their conclusions. \\

First, taking the same basis and functional as Kim \textit{et al.}, we find comparable results for the stable geometries of the \textit{p}-DAPA-CF$_3$ dual-fluorescent compound, as shown in Table~\ref{chap_06:table456}. The small difference is probably due to the presence of implicit solvent in Ref.~\cite{kim_singlebenzene_2023}. While calculations for the compounds \textit{p}-DAPA-C$_2$F$_5$ and \textit{p}-DAPA-C$_3$F$_7$ were not reported in the original paper, calculations performed in this work probed their PES and values are reported in Table~\ref{chap_06:table456}. Furthermore, the energies of the same type of geometries are probed for the newly proposed system \textit{p}-DAPA-CN. \\

From these energies, energetic barriers of the S$_1$ surface (Table~\ref{chap_06:table456_barrier}) and energetic gaps between S$_0$ and S$_1$, corresponding to absorption and emission wavelengths, can be calculated (Table~\ref{chap_06:talbe_S1S0gaps}). 
The calculations reported in the literature provide energy gaps between $S_1$ and $S_0$ close to the two fluorescence wavelengths measured experimentally, which inspires trust in these results (values are reported in the first two lines of Table~\ref{chap_06:talbe_S1S0gaps}) \cite{kim_singlebenzene_2023}. \\

   \begin{table}[bht!]
\caption{Energies in eV calculated at the B3LYP-D3/6-31++G** level. Energies are relative to the optimized ground state geometry $S_0$. The first value line corresponds to values reported in Ref.~\cite{kim_singlebenzene_2023} with implicit solvent.}
\label{chap_06:table456}
\begin{minipage}{1.0\textwidth}
\centering
\begin{tabular}{|c|c|c|c|c|c|c|}
\hline
{System / Geometry} & \boldmath$S_{1,FC}$ & \boldmath$S_{1,N^*}$ & \boldmath$S_{1,TS1}$ & \boldmath$S_{1,T^*}$  & \boldmath$S_{1,TS2}$& \boldmath$S_{1,TT}$  \\
\hline \textbf{\textit{p}-DAPA-CF$_3$} (\cite{kim_singlebenzene_2023}) & $3.12$ & $2.88$ & $2.98$ & $2.94$ & $3.10$ & $3.07$ \\
\hline \textbf{\textit{p}-DAPA-CF$_3$}& $3.141$ & $2.894$ & $2.997$ & $2.975$ & $3.116$ & $3.099$ \\
\hline \textbf{\textit{p}-DAPA-C$_2$F$_5$} & $3.138$ & $2.891$ & $2.991$ & $2.965$ & $3.106$ & $3.087$\\
\hline \textbf{\textit{p}-DAPA-C$_3$F$_7$} & $3.131$ & $2.884$ & $2.980$ & $2.956$ & $3.092$ & $3.067$\\
\hline \textbf{\textit{p}-DAPA-CN}& $3.103$ & $2.864$ & $2.962$ & $2.935$ & $3.067$ & $3.047$\\
\hline
\end{tabular}

\end{minipage}

\end{table}

   \begin{table}[bht!] 
\caption{Barrier energies in eV calculated at the B3LYP-D3/6-31++G** level. The first value line corresponds to values reported in Ref.~\cite{kim_singlebenzene_2023}.}
\label{chap_06:table456_barrier}
\begin{minipage}{1.0\textwidth}
\centering
\hspace*{-1.6cm} \begin{tabular}{|c|c|c|c|c|c|c|}
\hline
\, & \boldmath$E(S_{1,T^*}) - E(S_{1,N^*})$ & \boldmath$E(S_{1,TS1}) - E(S_{1,N^*})$ & \boldmath$E(S_{1,TS1}) - E(S_{1,T^*})$ \\
\hline \textbf{\textit{p}-DAPA-CF$_3$} (\cite{kim_singlebenzene_2023}) & $0.060$ & $0.10$  & $ 0.040$ \\
\hline \textbf{\textit{p}-DAPA-CF$_3$}& $0.0810$  & $0.103$  & $0.0221$ \\
\hline \textbf{\textit{p}-DAPA-C$_2$F$_5$} & $0.0738$  & $0.101$  & $0.0267$ \\
\hline \textbf{\textit{p}-DAPA-C$_3$F$_7$} & $0.0714$ & $0.0956$  & $0.0242$ \\
\hline \textbf{\textit{p}-DAPA-CN} & $0.0705$  & $0.0974$  & $0.0269$\\
\hline
\end{tabular}

\end{minipage}

\end{table}

       \begin{table}[bht!]
\caption{S$_1$-S$_0$ transition energies in eV (in nm in parenthesis) for the given geometry. FC corresponds to the Franck-Condon geometry which is the S$_0$ optimized geometry without proton transfer. The FC gap reflects the absorption wavelength. Experimental values (Exp.) are taken according to Ref.~\cite{kim_singlebenzene_2023}.}
\label{chap_06:talbe_S1S0gaps}
\begin{minipage}{1.0\textwidth}
\centering
\begin{tabular}{|c|c|c|c|c|c|c|}
\hline
\, & \textbf{FC} &  \textbf{N$^*$} & \textbf{T$^*$}  & \textbf{TT$^*$}\\
\hline \textbf{\textit{p}-DAPA-CF\boldmath$_3$} \textbf{Exp.} & (380) &  (466)  &  (584)  &  Not observed \\
\hline \textbf{\textit{p}-DAPA-CF\boldmath$_3$} (\cite{kim_singlebenzene_2023})& $3.12$ (398) & $2.61$ (476)& $2.12$ (585)  & $1.58$ (785) \\
\hline \textbf{\textit{p}-DAPA-CF\boldmath$_3$} & 3.141 (395) & 2.621 (473) & 2.108 (588) & 1.509 (822) \\
\hline \, & \, & \, & \, & \, \\
\hline \textbf{\textit{p}-DAPA-C\boldmath$_2$F\boldmath$_5$} \textbf{Exp.} &  (381) & (465)  & (583)  &   Not observed \\
\hline \textbf{\textit{p}-DAPA-C\boldmath$_2$F\boldmath$_5$} & 3.138 (395) & 2.632 (471) & 2.108 (588)  & 1.518 (817) \\
\hline \, & \, & \, & \, & \, \\
\hline \textbf{\textit{p}-DAPA-C\boldmath$_3$F\boldmath$_7$} \textbf{Exp.}  &  (381) & (463)  & (584)  &   Not observed \\
\hline \textbf{\textit{p}-DAPA-C\boldmath$_3$F\boldmath$_7$}  & 3.131 (396) & 2.629 (472) & 2.107 (588) & 1.511 (820)  \\
\hline \, & \, & \, & \, & \, \\
\hline \textbf{\textit{p}-DAPA-CN}  & 3.103 (400)  & 2.616 (474) & 2.103 (590) & 1.516 (818) \\
\hline
\end{tabular}

\end{minipage}

\end{table}

These results illustrate several main points. To begin with, ab initio S$_1$-S$_0$ energy gaps are close to the experimental values obtained with the absorption and emission spectra. Then, the absence of solvent effect in our calculations does not affect drastically the calculation results. Concerning the new system \textit{p}-DAPA-CN, we see that the barrier to reach T$^*$ is significantly reduced in comparison with \textit{p}-DAPA-CF$_3$ while T$^*$ is also stabilized with a lower energy compared to N$^*$ (Table~\ref{chap_06:table456_barrier}). Moreover, this substituent modification does not cause a significant shift in the emitted wavelengths (Table~\ref{chap_06:talbe_S1S0gaps}). \\
Finally, the influence of electron withdrawing substituents might be highlighted when analysing the calculated energies. Although the difference is small, the more electron-withdrawing effect carried by the substituent, the smaller the barrier from the N$^*$ to the T$^*$ states ($E(S_{1,TS1}) - E(S_{1,N^*})$). Therefore, regarding the ultrafast dynamics, the electron-withdrawing effect could potentially increase the probability to populate the T$^*$ state. \\
\subsection*{Transition state theory calculation}
From transition state theory, it is possible to estimate the transfer rate from the $S_{1,N^*}$ geometry to the $S_{1,T^*}$ geometry, assuming a purely thermally activated process and ignoring the ultrafast effects that might arise from the fact that the dynamics starts in the Franck-Condon state. In this context, one can approximate the transition rate as $k = \nu \exp(-\Delta E / k_B T)$ with $\Delta E$ the energy barrier and $\nu$ the frequency of the reaction coordinate in the reactant state. As the precise value of $\nu$ is unknown, we can employ different approaches: in the extreme case, $\nu$ is purely determined by the hydrogen stretching frequency ($\nu= \frac{1}{10}~$fs$^{-1}$) and one obtains $\tau = (1/k) \approx 500~$fs at 300K and $\tau \approx 30~$fs at 1000K. This value of $\nu$ is probably overestimated as slower structural deformations of the molecule, affecting the donor-acceptor distance might certainly contribute to the proton transfer. Indeed the corresponding rate values are higher than that estimated from the molecular dynamics simulations (see main text). More accurate rates can be obtained considering a value of frequency closer to the transition state, $\nu = 1/30 ~$fs$^{-1}$ which gives  $\tau  \approx 1600$~fs at 300K and $\tau  \approx 100$~fs at 1000K. \\

An other approach is to use Eyring's formula, assuming that the free energy $\Delta^{\ddag} G^{\circ}$ is well defined in the excited state, one can also estimate the rate from 
\begin{equation}
    k(T) = \frac{k_B T}{hc^{\circ}}e^{-\Delta^{\ddag} G^{\circ}/RT}
\end{equation} with $c^{\circ}$ the concentration (taken here $c^{\circ}=1$) and $R$ the gas constant. We here take the free energy as being equal to $\Delta E$, neglecting the difference in entropy between N$^*$ and T$^*$. From this, one obtains for \textit{p}-DAPA-CF$_3$, $\tau \approx 8600$~fs at 300K and $\tau \approx 159$~fs at 1000K. For \textit{p}-DAPA-CN, one obtains $\tau \approx 6900$~fs  at 300K and  $\tau \approx 149$~fs at 1000K. These transfer lifetimes calculated at 1000K are in the same range than the molecular dynamics results, indicating a thermally driven process.

\subsection*{Side note on the study of the emission spectrum}

\begin{figure}[bht!]
    \centering
    \includegraphics[scale=0.43]{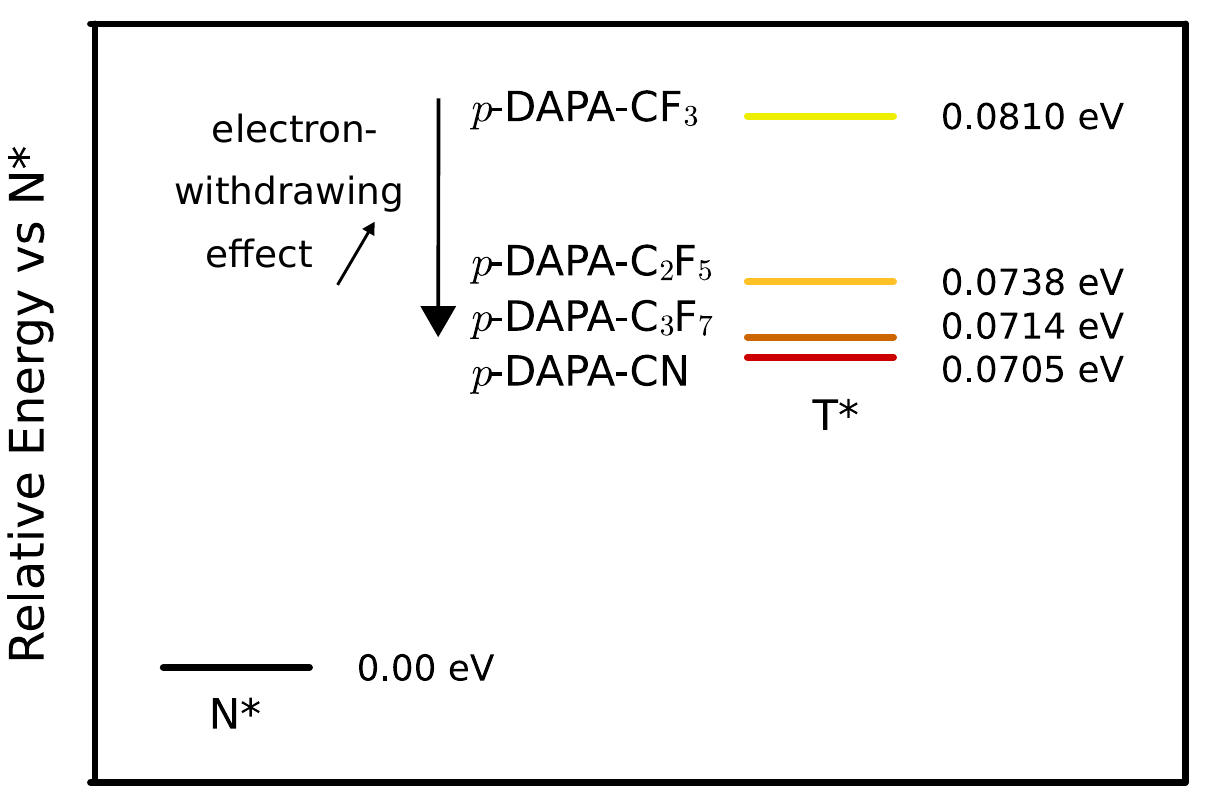}
    \caption{Calculated energies of the T$^*$ state relative to N$^*$ at B3LYP-D3/6-31++G** level. The increase of the electron withdrawing effect lowers the T$^*$ energy which is in a consistent trend with a more intense T$^*$ peak for a thermal distribution, as suggested in Ref.~\cite{kim_singlebenzene_2023}.}
    \label{fig_chap06:exp_vs_S1_boltzmann}
\end{figure}

With regard to a possible "thermal distribution", the substituents also modify the ratio of energy between the N$^*$ and the T$^*$ states ($E(S_{1,T^*}) - E(S_{1,N^*})$ in Table~\ref{chap_06:table456_barrier}. This can be directly linked to the experimental emission spectra, as illustrated in Fig.~\ref{fig_chap06:exp_vs_S1_boltzmann}. From a thermal distribution, the more stable the state, the more populated it becomes and the higher the corresponding peak in the fluorescence spectrum. Experimentally, Kim \textit{et al.} found a second peak at higher wavelength, corresponding to the T$^*$ state with an intensity that increases with the number of fluorines. The calculated $E(S_{1,T^*})$ goes in this direction, stabilizing the transferred state with more and more F atoms. In this same logic, the new system \textit{p}-DAPA-CN is expected to have a higher T$^*$ peak than the highest measured experimentally. However, calculations do not explain the reason why the T$^*$ is higher than the N$^*$ emission in the measured spectra~\cite{kim_singlebenzene_2023}. Although the two proton-donors and two proton-acceptors can lead to two distinct single proton transfers, leading to two different T$^*$ states for the same system, taking this factor two into account for a thermal distribution still underestimates the T$^*$ peak intensity by one order of magnitude. The electronic dipole between S$_0$ and S$_1$, which enters the emission probability, is here not taken into account. Since numerical results are very sensitive to the ab initio values, more accurate TD-DFT parameters may give more intuitions on the discrepancy between the experimental and the numerical ratio.

\section{Molecular Dynamics parameters and results}

\subsection*{Dynamics Parameters and Ground State Sampling}

For Born-Oppenheimer molecular dynamics, initial conditions chosen to propagate dynamics are extracted from ground state sampling. In order to do so, several dynamics of 1 picosecond with a time step of $d t = 1.0~$fs in the S$_0$ surface are propagated in the NVT ensemble using a Langevin thermostat with a damping rate of $0.1~$fs$^{-1}$. Coordinates and velocities are picked every 200 fs from these dynamics as initial conditions for the dynamics onto the excited state S$_1$. Ground state sampling is performed at 300 Kelvin and 1000 Kelvin. \\
\indent Molecular dynamics on S$_1$ are then carried out with a time step of $d t = 1.0~$fs in the NVE ensemble. The set of trajectories include 10 trajectories at 300 K for $p$-DAPA-CF$_3$ and for $p$-DAPA-CN, 14 at 1000 K for $p$-DAPA-CF$_3$ and for $p$-DAPA-CN. These dynamics are carried out for a simulated time of 1 ps except if a proton transfer occurs within this first picosecond of simulation.

\subsection*{Trajectory details}

We give here some details about the different molecular dynamics trajectories. First, the detail of the trajectories with the observed proton transfer time for reactive dynamics are presented in Table~\ref{chap06_appendix:table_dyn}.
      \begin{table}[ht]
\caption{Time simulated (fs) for different temperatures/systems (horizontal axis) and initial conditions (vertical axis). \checkmark[green] means a reaction occurred at the indicated time. \textcolor{red}{X} indicates the absence of proton transfer whereas \textcolor{teal}{$\times 2$} denotes a double proton transfer}
\label{chap06_appendix:table_dyn}

\begin{minipage}{1.0\textwidth}
\centering
\hspace*{-1.0cm} \begin{tabular}{|c|c|c|c|c|}
\hline
\,  & $300$K \textit{p}-DAPA-CF$_3$&  $300$K \textit{p}-DAPA-CN& $1000$K \textit{p}-DAPA-CF$_3$& $1000$K \textit{p}-DAPA-CN \\
\hline GS sampling 1 & \textcolor{red}{X} & \textcolor{red}{X} &  165 (\checkmark[green]) & 92 (\checkmark[green]) \\
\hline GS sampling 2 & \textcolor{red}{X} & \textcolor{red}{X} & 332 (\checkmark[green]) & 135 (\checkmark[green])   \\
\hline GS sampling 3 & \textcolor{red}{X} & \textcolor{red}{X} & 257 (\checkmark[green]) &  127 (\checkmark[green])  \\
\hline GS sampling 4 & \textcolor{red}{X} & \textcolor{red}{X} &  279 (\checkmark[green]) &  357 (\checkmark[green]) \\
\hline GS sampling 5 & \textcolor{red}{X} & 617 (\checkmark[green]) & 69 (\checkmark[green]) &  313 (\checkmark[green]) \\
\hline GS sampling 6 & \textcolor{red}{X} & \textcolor{red}{X} & 21 (\checkmark[green]) &   70  (\checkmark[green]) \\
\hline GS sampling 7 & \textcolor{red}{X} & \textcolor{red}{X} & 240 (\checkmark[green])& 42 (\checkmark[green]) \\
\hline GS sampling 8 & \textcolor{red}{X} & \textcolor{red}{X} & 493 (\checkmark[green]) &   38 (\checkmark[green])\\
\hline GS sampling 9 & \textcolor{red}{X} & \textcolor{red}{X} & \textcolor{red}{X} & 43 (\checkmark[green])  \\
\hline GS sampling 10 & \textcolor{red}{X} & \textcolor{red}{X} & 885 (\checkmark[green])&  312 (\checkmark[green])   \\
\hline GS sampling 11 &  &  & 446 (\checkmark[green]) & 144 (\checkmark[green])  \\
\hline GS sampling 12 & &  & 312 (\checkmark[green]) &  164 (\checkmark[green])  \\
\hline GS sampling 13 &  &  & 38 (\checkmark[green]) &  297 (\checkmark[green])  \\
\hline GS sampling 14 &  &  &  102 (\checkmark[green]) &  87 (\checkmark[green])  \\
\hline
\end{tabular}

\end{minipage}

\end{table}

From these values, the average transfer time is calculated. Taking into account the 13 reactive trajectories at 1000 K for \textit{p}-DAPA-CF$_3$ and the 14 reactive trajectories for \textit{p}-DAPA-CN, the mean values are $\Bar{X}$(\textit{p}-DAPA-CF$_3$) = 280 fs and $\Bar{X}$(\textit{p}-DAPA-CN) = 159 fs. Then, $S^2 = \frac{1}{n-1}\sum_{i=1}^{n} \left( X_i - \Bar{X}\right)^2$  is calculated to indicate the 95\% confidence interval. It gives $S^2$(\textit{p}-DAPA-CF$_3$) = 54990 fs$^2$ and $S^2$(\textit{p}-DAPA-CN) = 12815 fs$^2$. From the statistics table, we find the t value for \textit{p}-DAPA-CF$_3$ equals  2.179  and 2.160 for \textit{p}-DAPA-CN. Finally, $t \times \frac{S}{\sqrt(n)}$(\textit{p}-DAPA-CF$_3$) = 145 fs and  \quad $t \times \frac{S}{\sqrt(n)}$(\textit{p}-DAPA-CN)= 66 fs. A summary can be found in Table~\ref{chap_06:table_dynamics_300_1000}. It must be emphasized that the mean values of proton transfer are here calculated from the raw values directly. In the main text, the proton transfer is determined by fitting the proportion of reactive trajectories over time, which gives different but also consistent average transfer times. \\

\begin{table}[ht] 
\caption{Molecular dynamics results for \textit{p}-DAPA-CF$_3$ and \textit{p}-DAPA-CN. Dynamics are analysed for a simulated time of 1 ps. The average transfer time at 1000K is given within 95\% confidence.}
\label{chap_06:table_dynamics_300_1000}
\begin{minipage}{1.0\textwidth}
\centering
\begin{tabular}{|c|c|c|}
\hline
\, & \textbf{\textit{p}-DAPA-CF$_3$} &  \textbf{\textit{p}-DAPA-CN} \\
\hline \# Trajectory 300K & 10 & 10 \\
\hline \# Proton Transfer 300K  & 0 & 1 \\
\hline \# Trajectory 1000K & 14 & 14 \\
\hline \# Proton Transfer 1000K & 13 & 14 \\
\hline Average Transfer Time 1000K & 280 $\pm$ 145 fs &  159 $\pm$ 66 fs\\
\hline
\end{tabular}
\end{minipage}

\end{table}

\subsection*{Benzene Bond Dynamics}
\addcontentsline{toc}{subsection}{\protect\numberline{}\hspace*{-1.0cm}\small{Study of the Benzene Bonds in Time}}

We now focus on the molecular details during the dynamics. As a matter of fact, the benzene is thought as a major component of the system evolution. Since the studied systems are reduced in size, the conjugated ring, which is the central motif structure, might be able to drive the proton transfer or, at least, bear signature of its interplay with the reaction. \\
\begin{figure}[bht!]
    \centering
\hspace*{-1.65cm}  \includegraphics[scale=0.43]{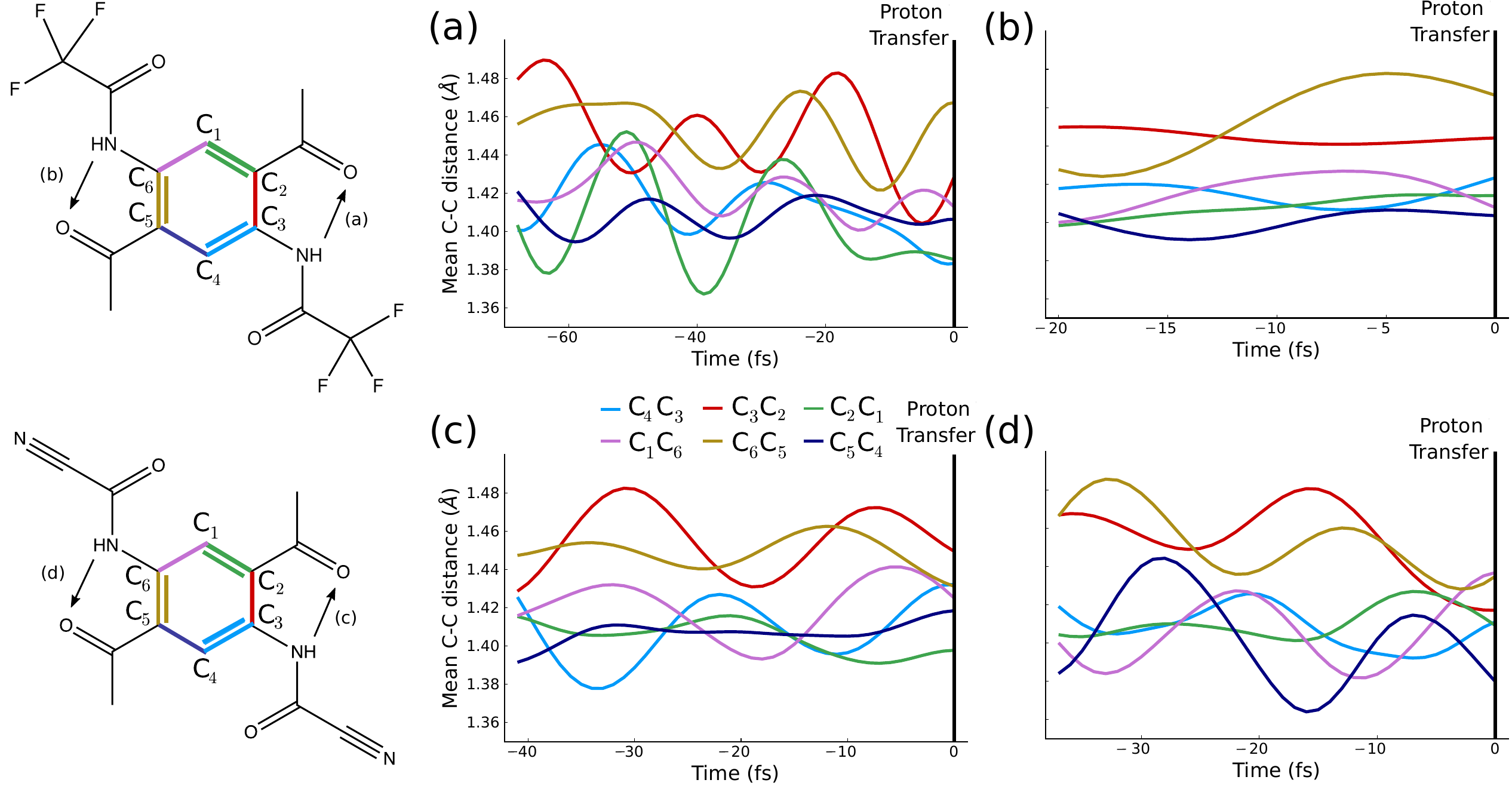}
    \caption{Mean benzene ring C-C distances averaged over the reactive dynamics at 1000 Kelvin both for (a)-(b) $p$-DAPA-CF$_3$ and (c)-(d) $p$-DAPA-CN. The proton transfer time is taken as reference and the average is represented for time preceding the transfer. The two possible proton transfers are differentiated as illustrated in the chemical structures and the average is performed for both proton transfers independently.}
    \label{fig_supmat:CF3_moyCC}
\end{figure}

A first step to analyse such hypothesis is to follow the ring carbon-carbon distances. Figure~\ref{fig_supmat:CF3_moyCC} shows the mean C-C distances averaged over the reactive dynamics at 1000 Kelvin both for $p$-DAPA-CF$_3$ and $p$-DAPA-CN. The proton transfer time is taken as the time reference and the average is done at reversed time (5~fs before proton transfer, 10~fs before proton transfer ...). Moreover, since two different single proton transfers are possible, the two types of dynamics are differentiated. The average is therefore calculated for both transfers, separating dynamics involving different protons. \\

Figure~\ref{fig_supmat:CF3_moyCC} shows a generally longer bond between carbons that bridge the proton-donor and the proton-acceptor. Whereas the lack of a clear signature can be attributed to the reduced sampling of reactive dynamics, this result highlights the difficulty in attributing a role to the benzene core concerning the proton transfer.  Therefore, we cannot conclude on the benzene effect on the proton transfer based solely on these observables. 

\subsection*{Adiabatic energetic gaps}
Here, we present the energy gap calculated during reactive dynamics at 1000 K for \textit{p}-DAPA-CF$_3$ and \textit{p}-DAPA-CN. In these dynamics, the ground state S$_0$ stays at more than 2.0~eV from the S$_1$ surface, which excludes non-adiabatic transitions between these two states. Concerning the second excited state S$_2$, it generally remains from 0.2~eV to 0.4~eV above S$_1$ for the first femtoseconds of the dynamics. Furthermore, the transition S$_0$ $\rightarrow$ S$_2$ has no oscillator strength, meaning that S$_2$ is not populated by the promoting pulse. In some cases, the S$_2$ surface momentarily comes closer to S$_1$, reaching values below 0.2~eV for a few fs in a few trajectories. Although these events seem unrelated to the proton transfer as they are well separated in time, this proximity between surfaces could justify a future semiclassical approach to probe potential non-adiabatic transitions between S$_1$ and S$_2$. 

\begin{figure}[bht!]
     \centering
     \begin{subfigure}[b]{0.3\textwidth}
         \centering
         \includegraphics[width=1.0\textwidth]{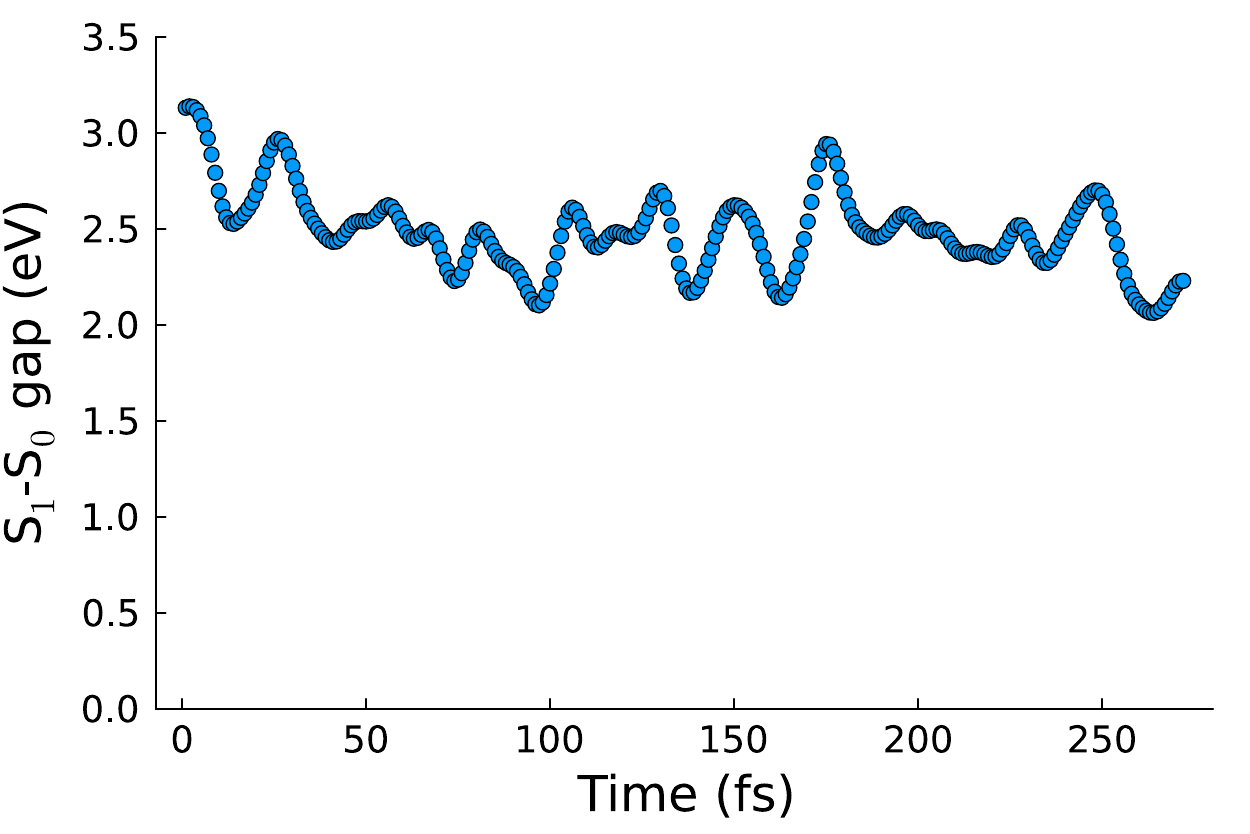}
     \end{subfigure}
     \hfill
     \begin{subfigure}[b]{0.3\textwidth}
         \centering
         \includegraphics[width=1.0\textwidth]{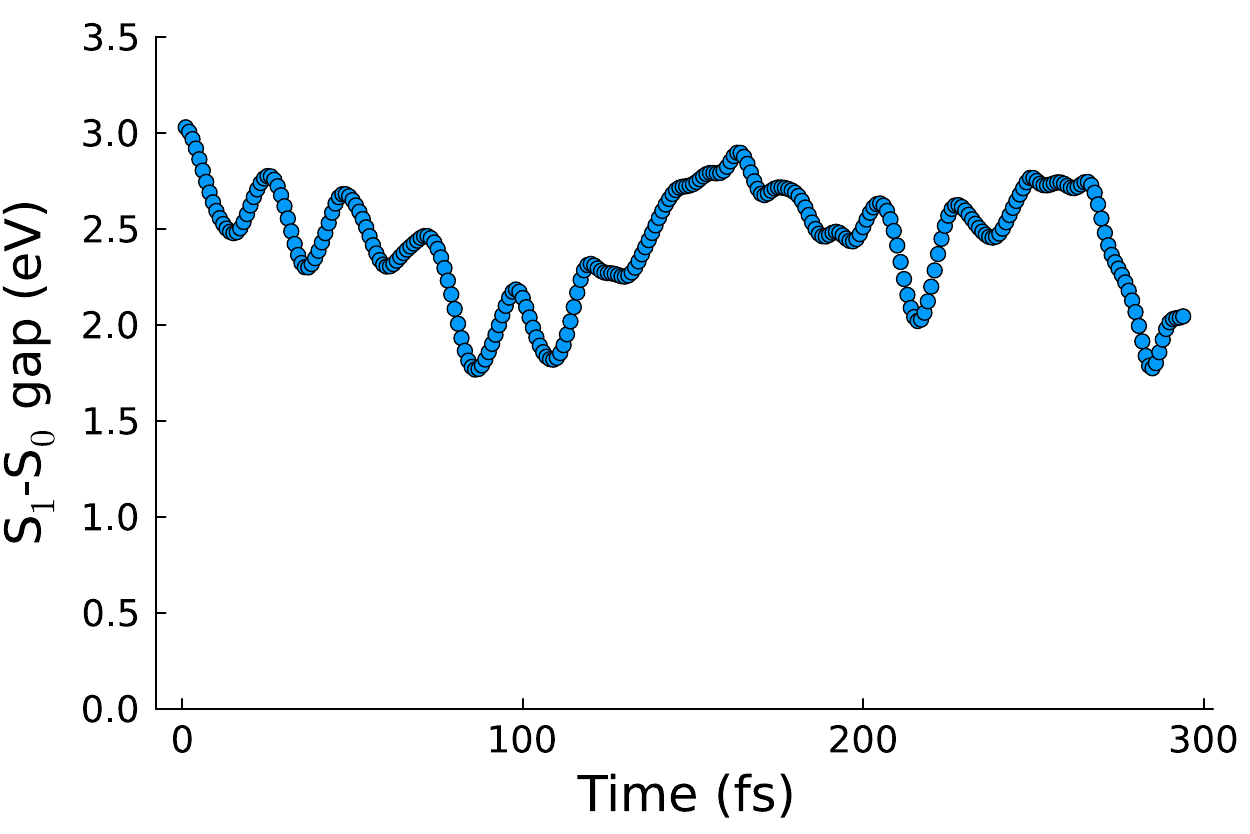}
     \end{subfigure}
     \hfill
        \begin{subfigure}[b]{0.3\textwidth}
         \centering
         \includegraphics[width=1.0\textwidth]{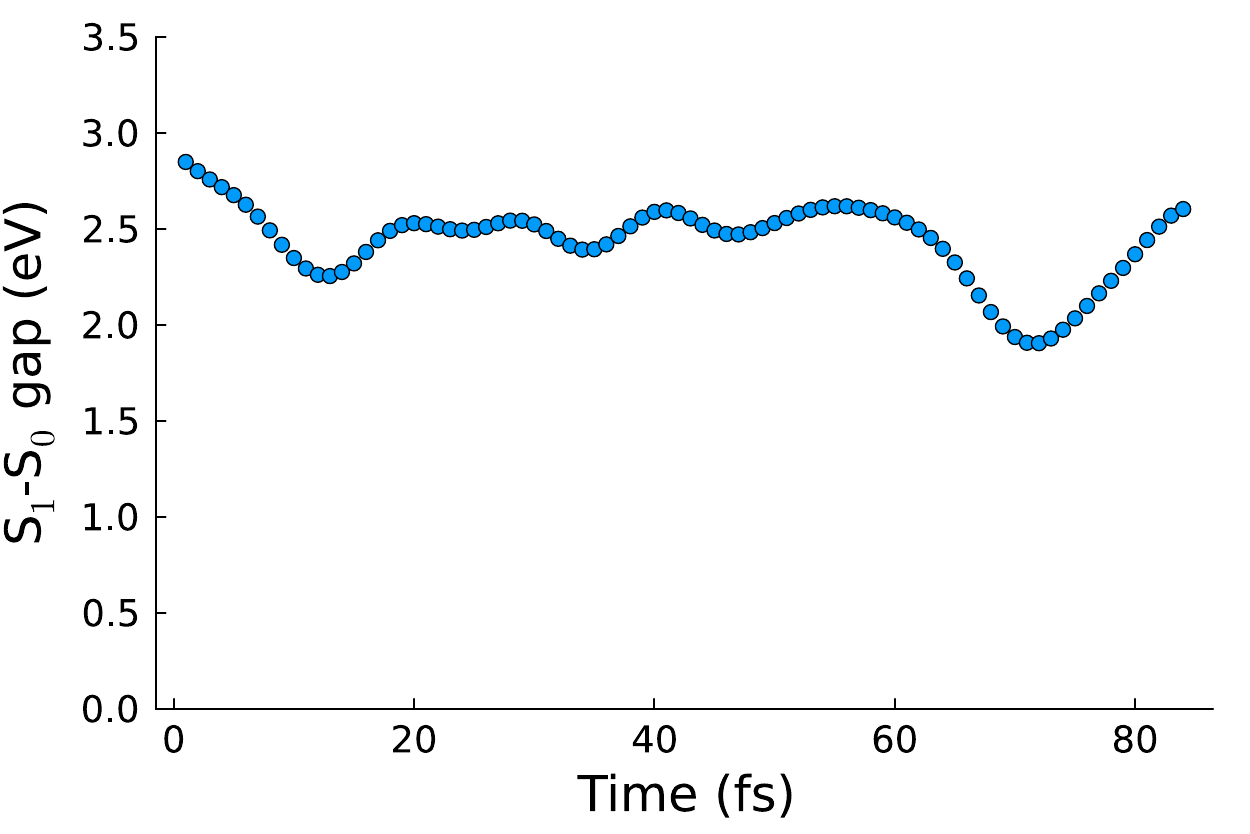}
     \end{subfigure}
     \hfill
        \begin{subfigure}[b]{0.3\textwidth}
         \centering
         \includegraphics[width=1.0\textwidth]{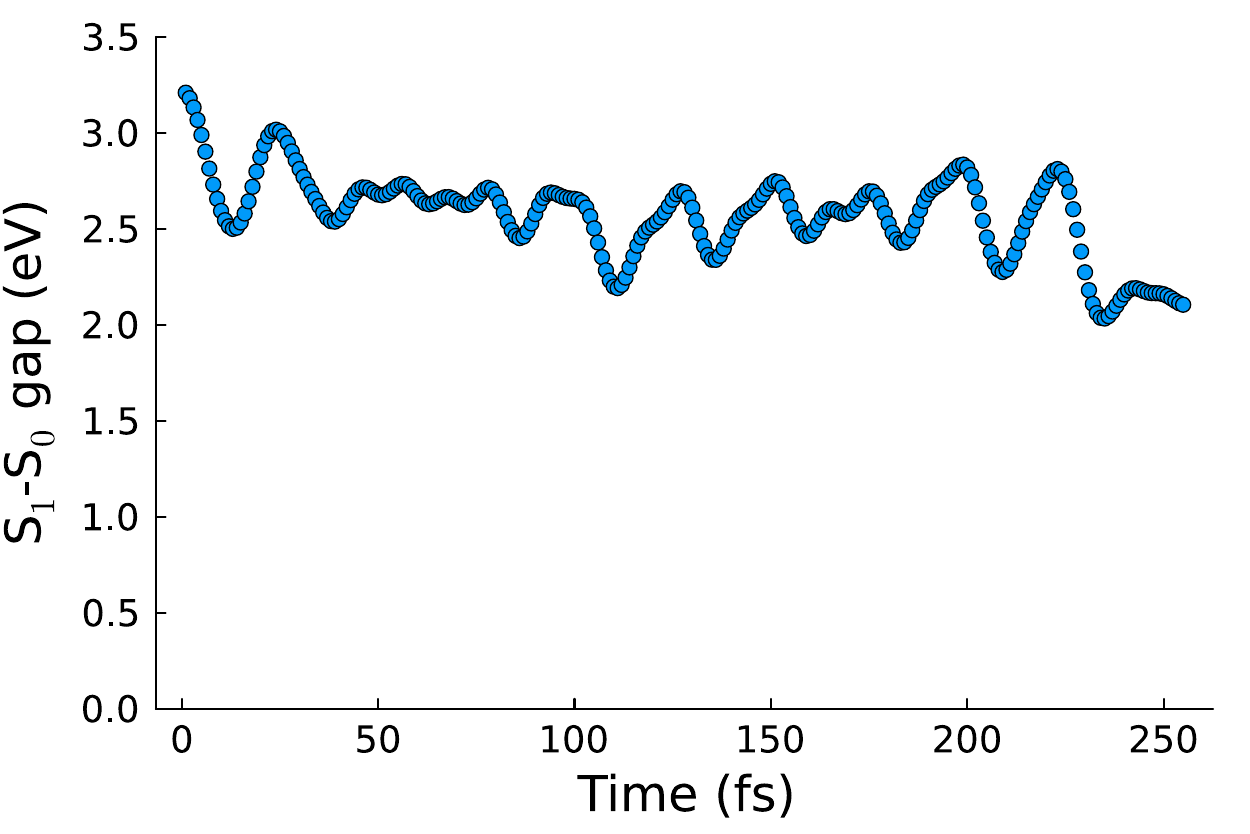}
     \end{subfigure}
     \hfill
        \begin{subfigure}[b]{0.3\textwidth}
         \centering
         \includegraphics[width=1.0\textwidth]{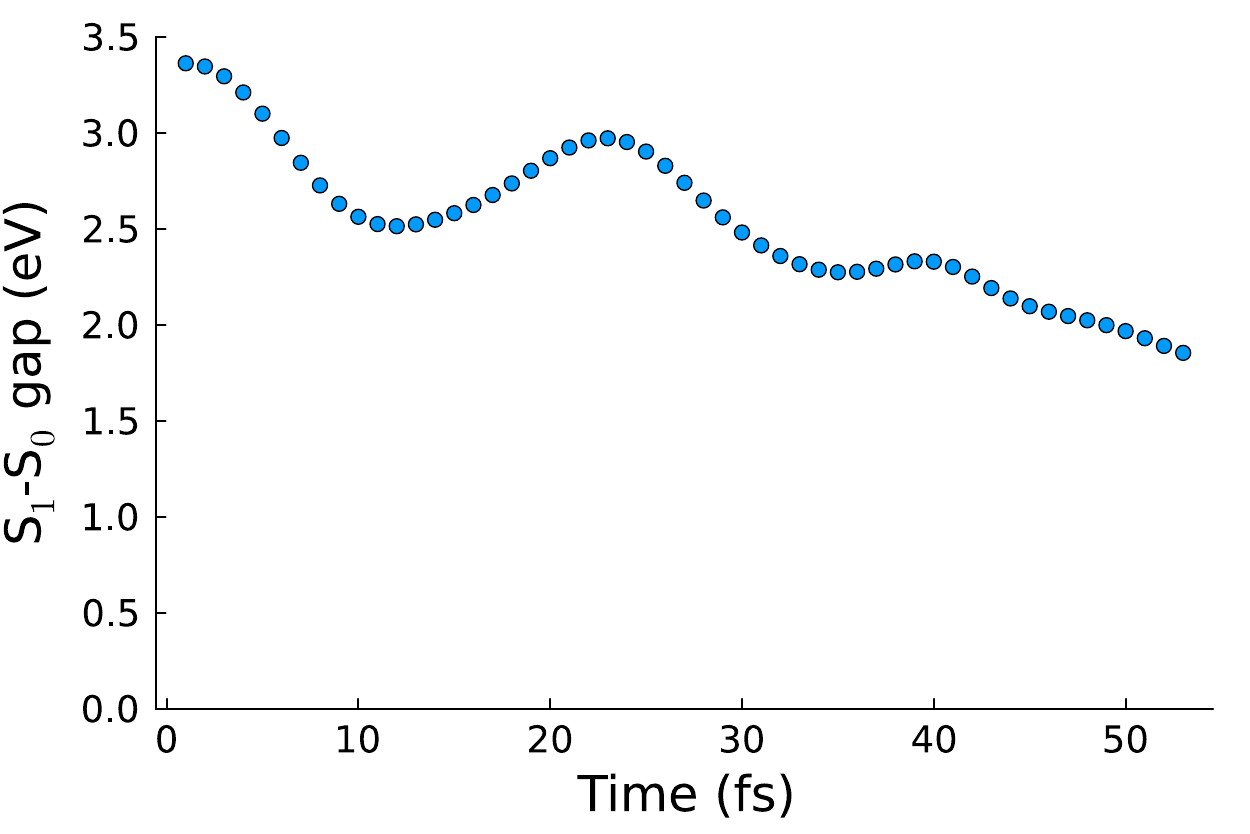}
     \end{subfigure}
        \hfill
        \begin{subfigure}[b]{0.3\textwidth}
         \centering
         \includegraphics[width=1.0\textwidth]{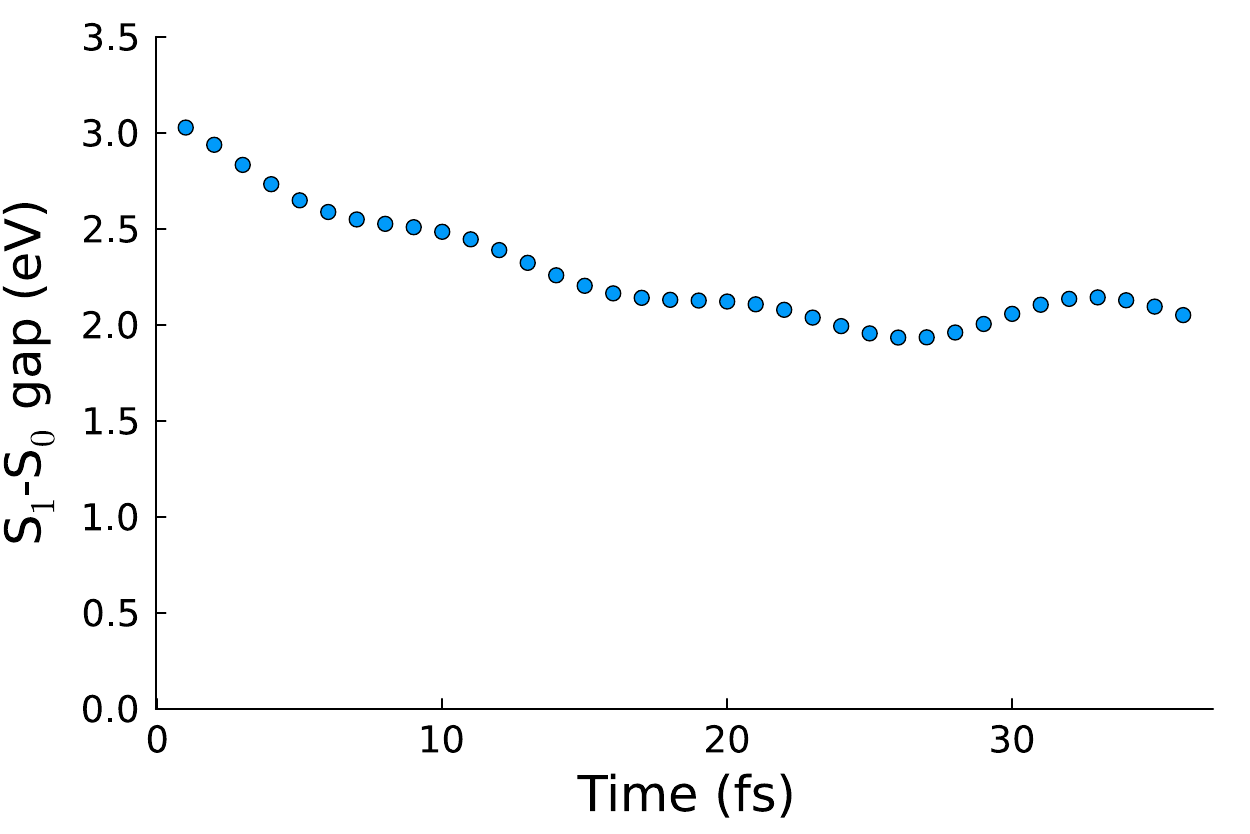}
     \end{subfigure}
        \caption{Energy gap between S$_0$ and S$_1$ for reactive dynamics at 1000 K of \textit{p}-DAPA-CF$_3$. The final time shown is the proton transfer time.}
        \label{fig_chap06:GapS1S0CF3}
\end{figure}

\begin{figure}[bht!]
     \centering
     \begin{subfigure}[b]{0.3\textwidth}
         \centering
         \includegraphics[width=1.0\textwidth]{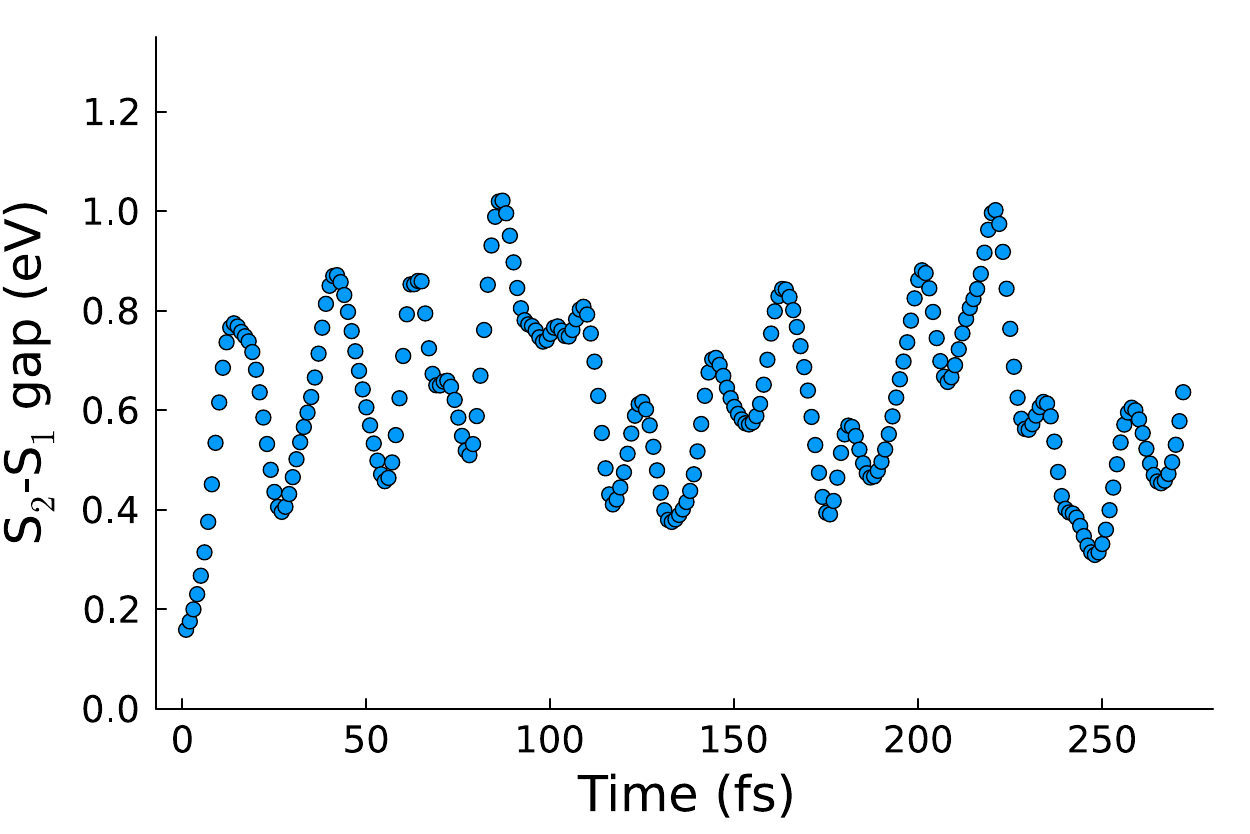}
     \end{subfigure}
     \hfill
     \begin{subfigure}[b]{0.3\textwidth}
         \centering
         \includegraphics[width=1.0\textwidth]{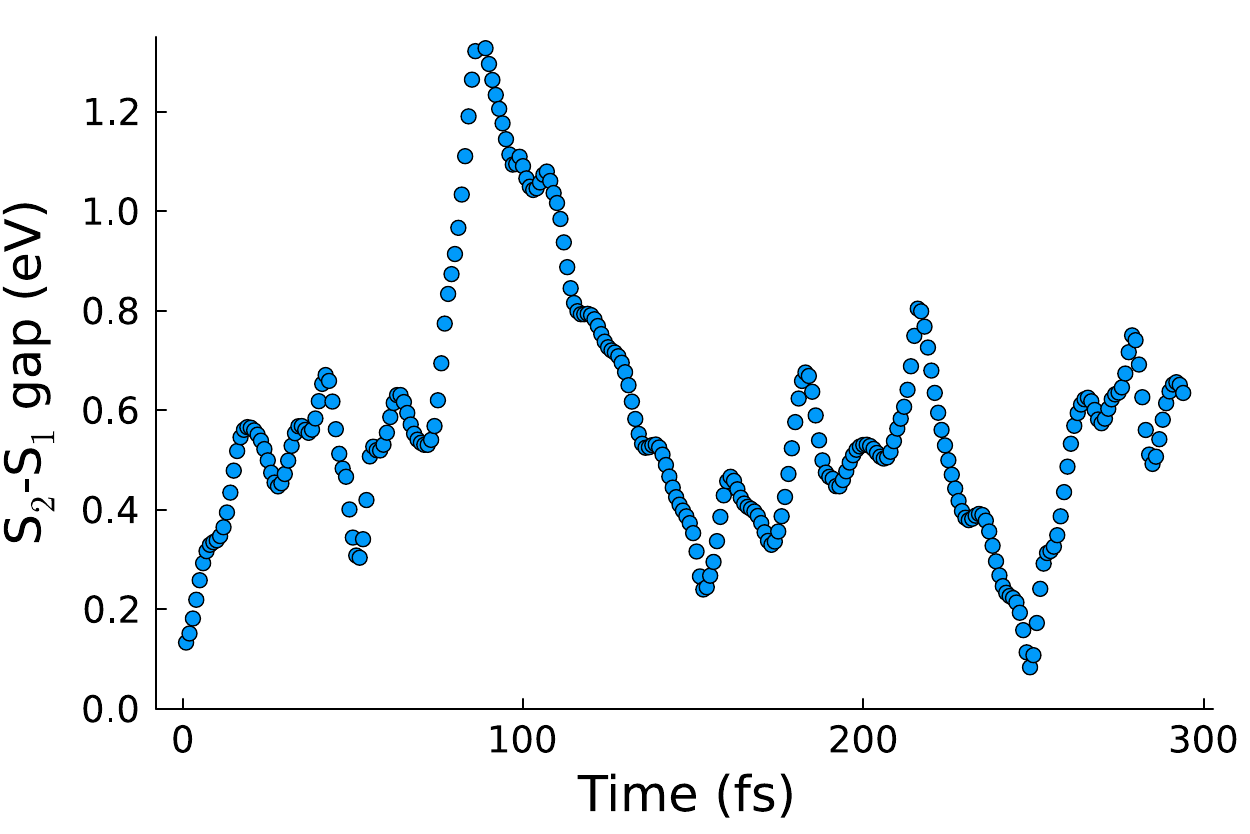}
     \end{subfigure}
     \hfill
        \begin{subfigure}[b]{0.3\textwidth}
         \centering
         \includegraphics[width=1.0\textwidth]{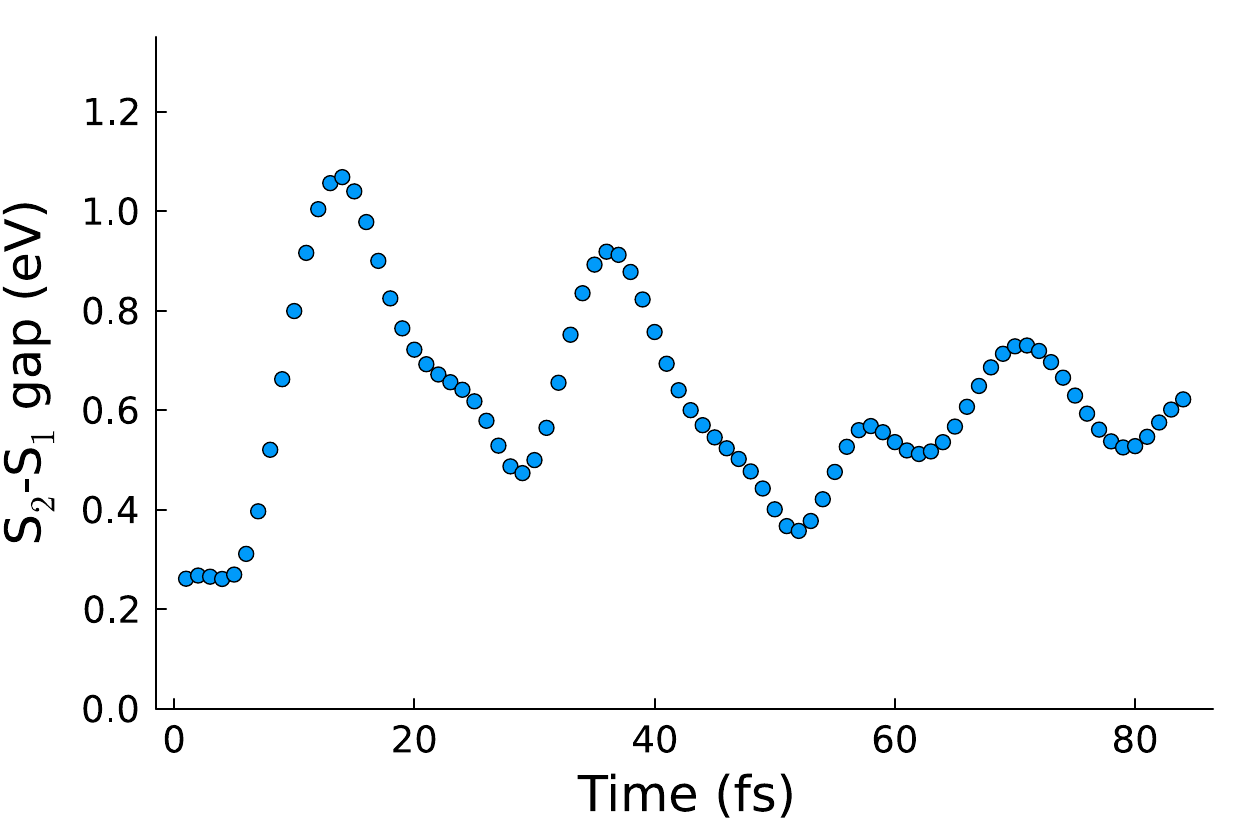}
     \end{subfigure}
     \hfill
        \begin{subfigure}[b]{0.3\textwidth}
         \centering
         \includegraphics[width=1.0\textwidth]{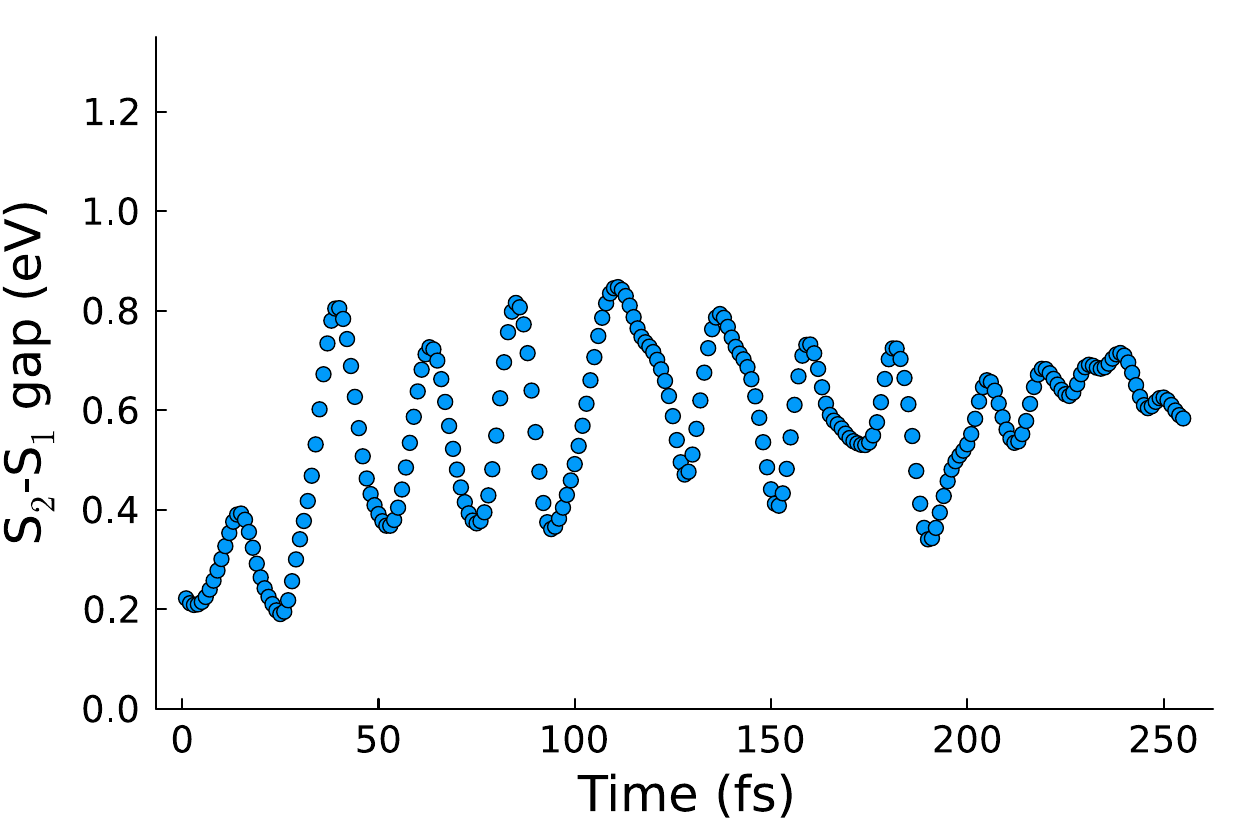}
     \end{subfigure}
     \hfill
        \begin{subfigure}[b]{0.3\textwidth}
         \centering
         \includegraphics[width=1.0\textwidth]{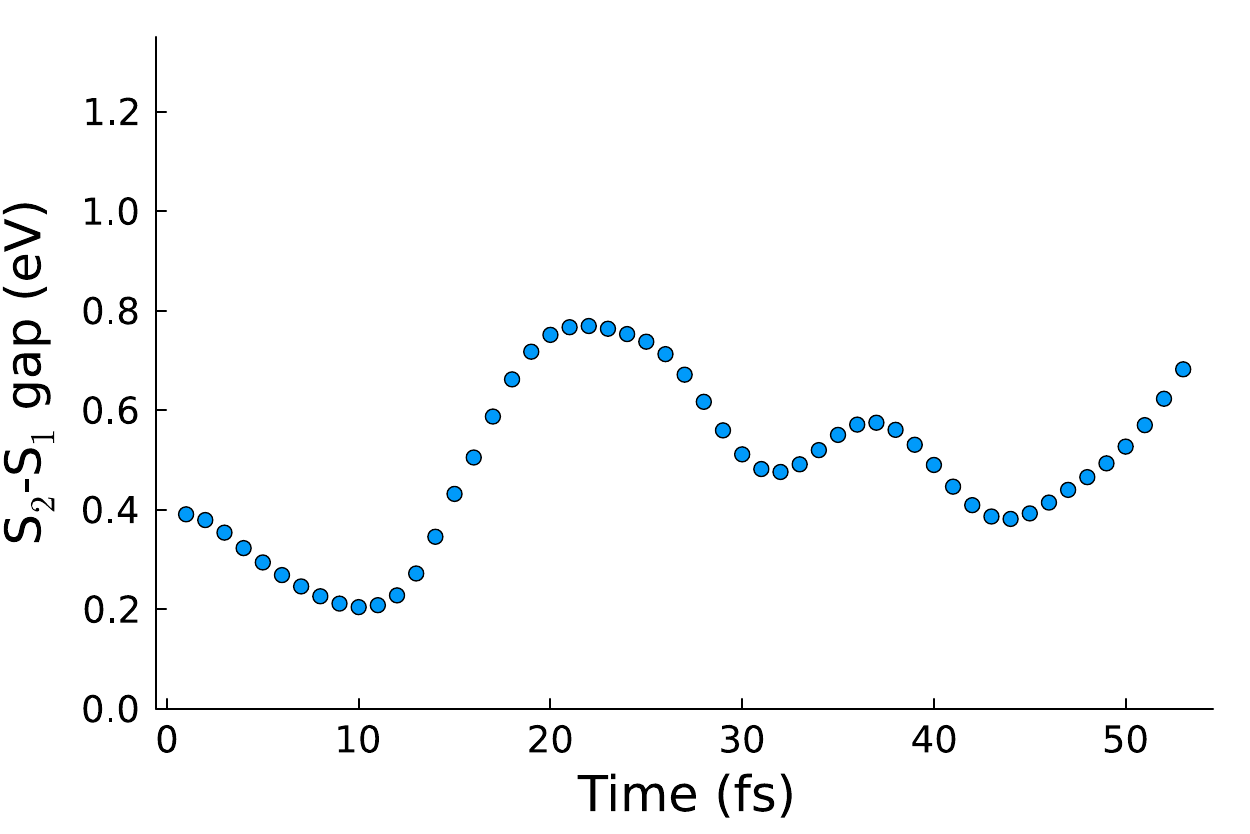}
     \end{subfigure}
        \hfill
        \begin{subfigure}[b]{0.3\textwidth}
         \centering
         \includegraphics[width=1.0\textwidth]{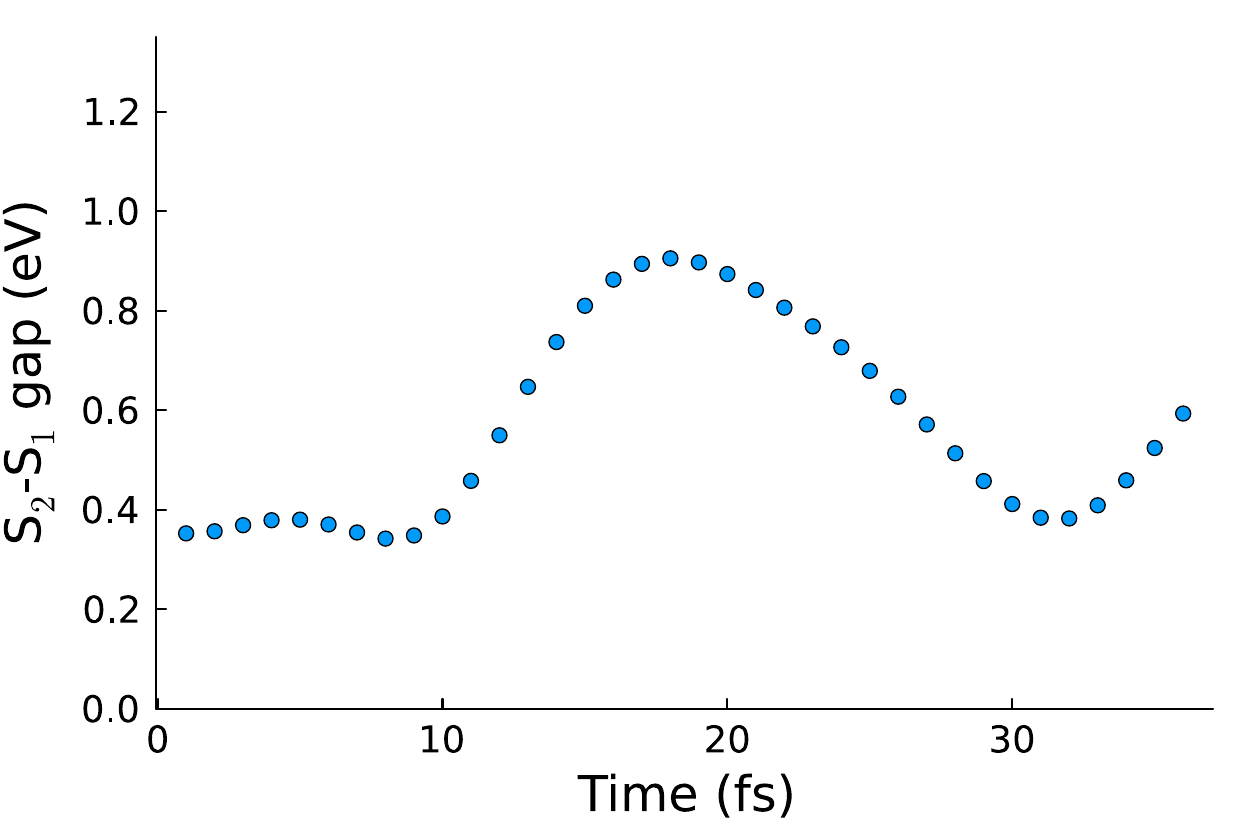}
     \end{subfigure}
         \caption{Energy gap between S$_2$ and S$_1$ for reactive dynamics at 1000 K of \textit{p}-DAPA-CF$_3$. The final time shown is the proton transfer time.}
        \label{fig_chap06:GapS2S1CF3}
\end{figure}

\begin{figure}[bht!]
     \centering
     \begin{subfigure}[b]{0.3\textwidth}
         \centering
         \includegraphics[width=1.0\textwidth]{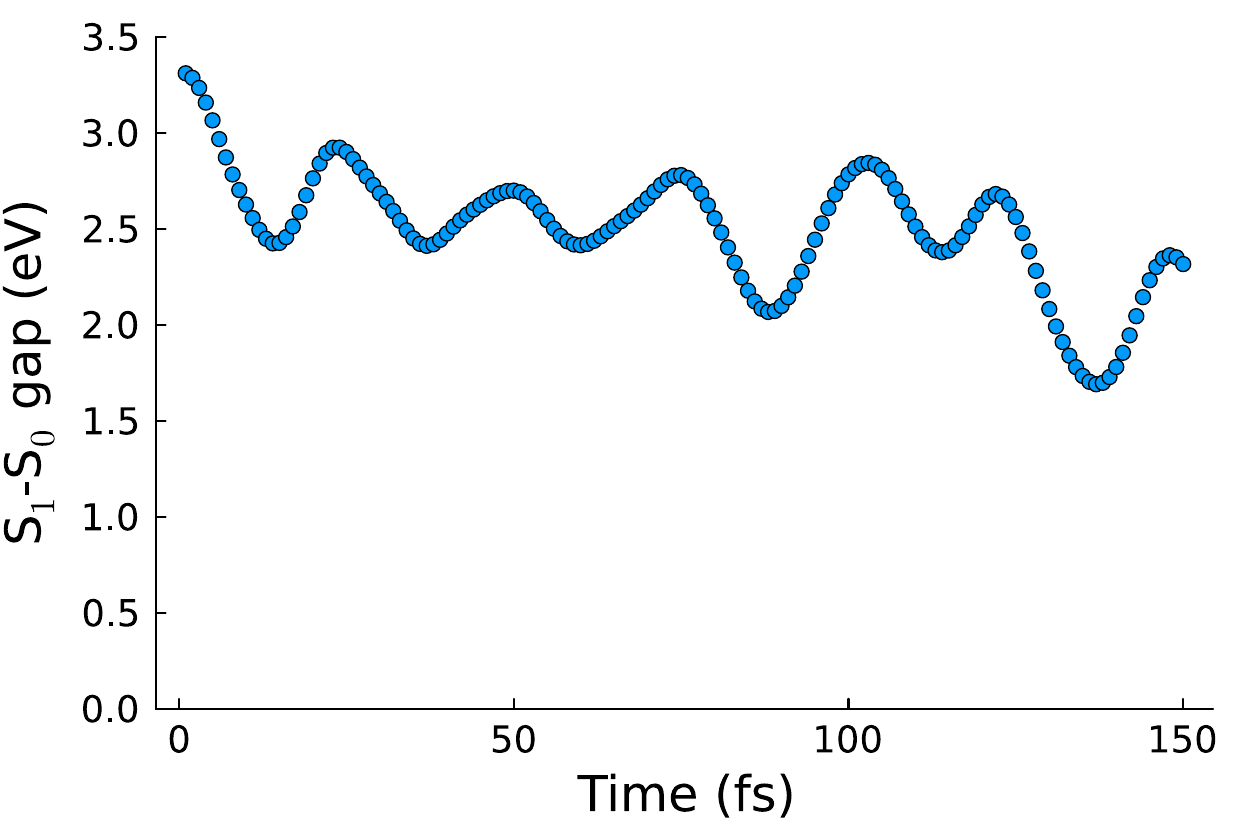}
     \end{subfigure}
     \hfill
     \begin{subfigure}[b]{0.3\textwidth}
         \centering
         \includegraphics[width=1.0\textwidth]{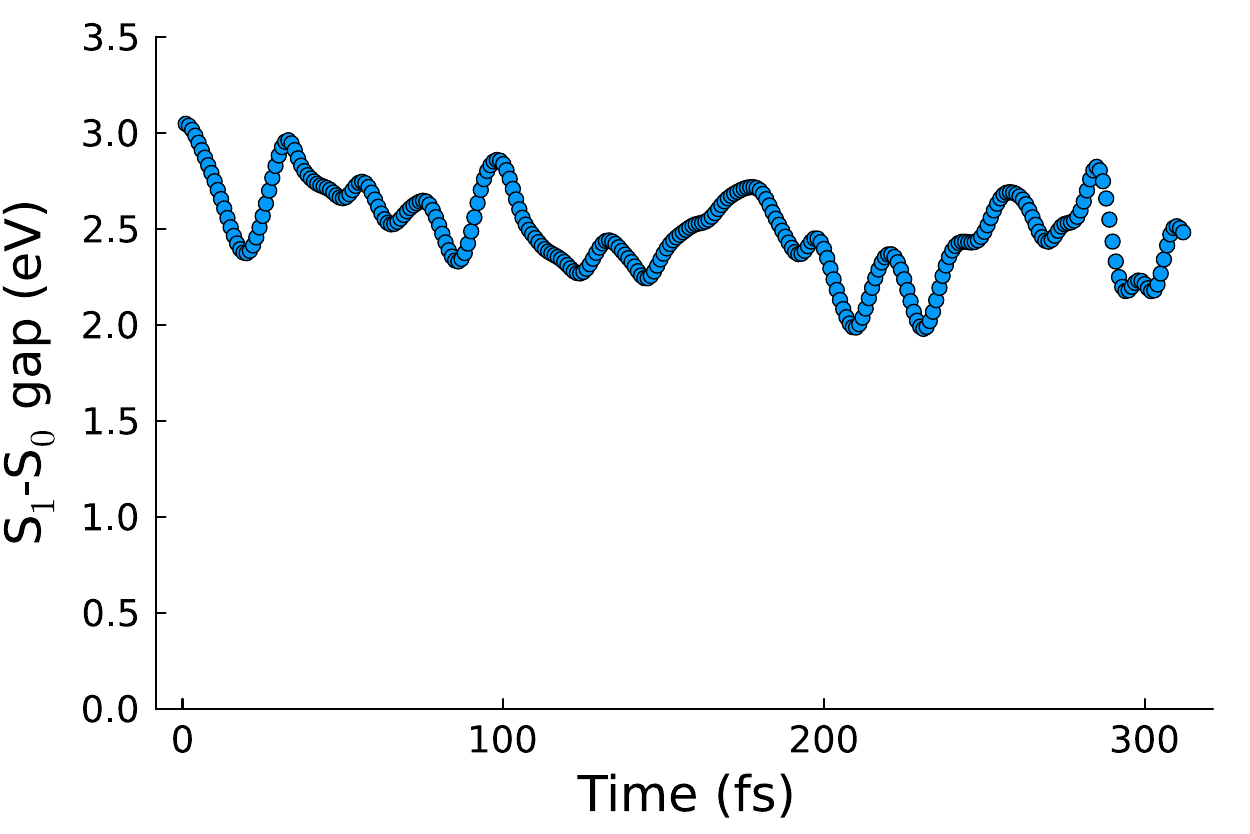}
     \end{subfigure}
     \hfill
        \begin{subfigure}[b]{0.3\textwidth}
         \centering
         \includegraphics[width=1.0\textwidth]{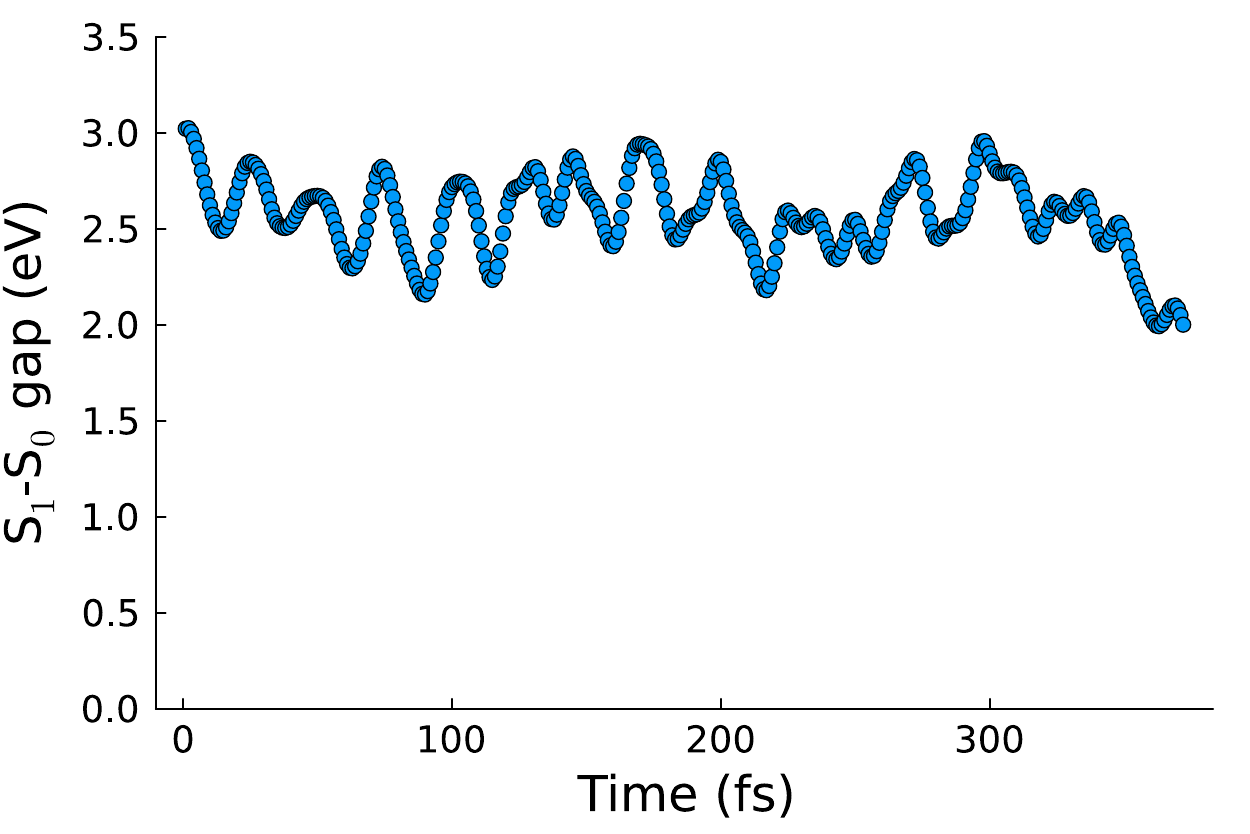}
     \end{subfigure}
     \hfill
        \begin{subfigure}[b]{0.3\textwidth}
         \centering
         \includegraphics[width=1.0\textwidth]{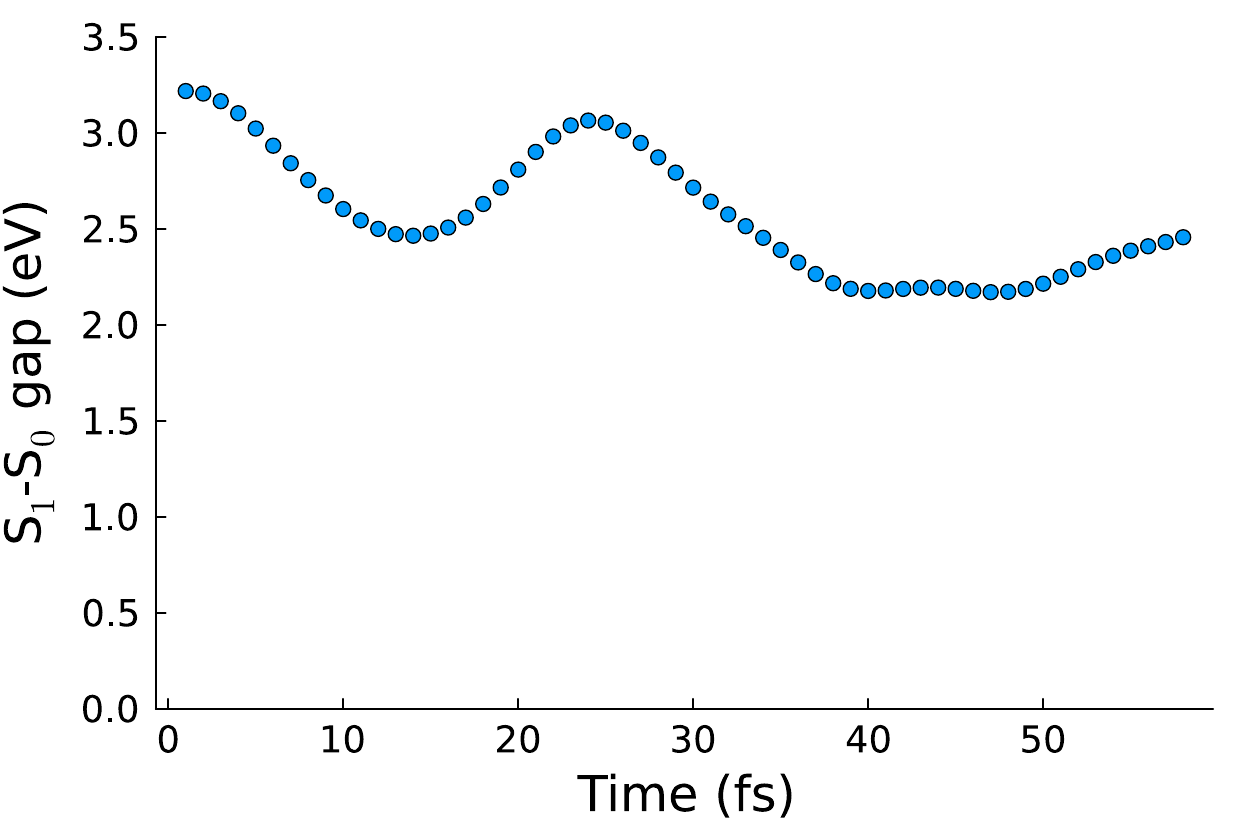}
     \end{subfigure}
     \hfill
        \begin{subfigure}[b]{0.3\textwidth}
         \centering
         \includegraphics[width=1.0\textwidth]{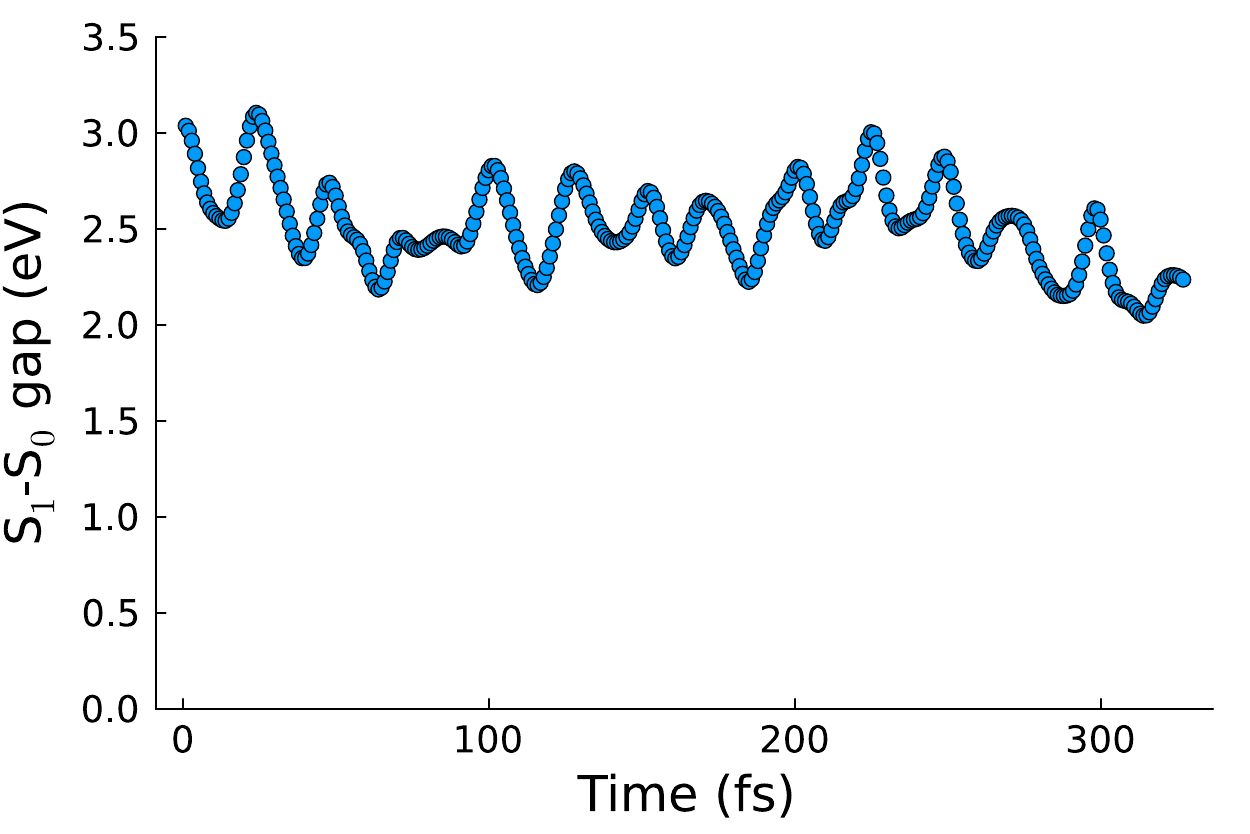}
     \end{subfigure}
        \hfill
        \begin{subfigure}[b]{0.3\textwidth}
         \centering
         \includegraphics[width=1.0\textwidth]{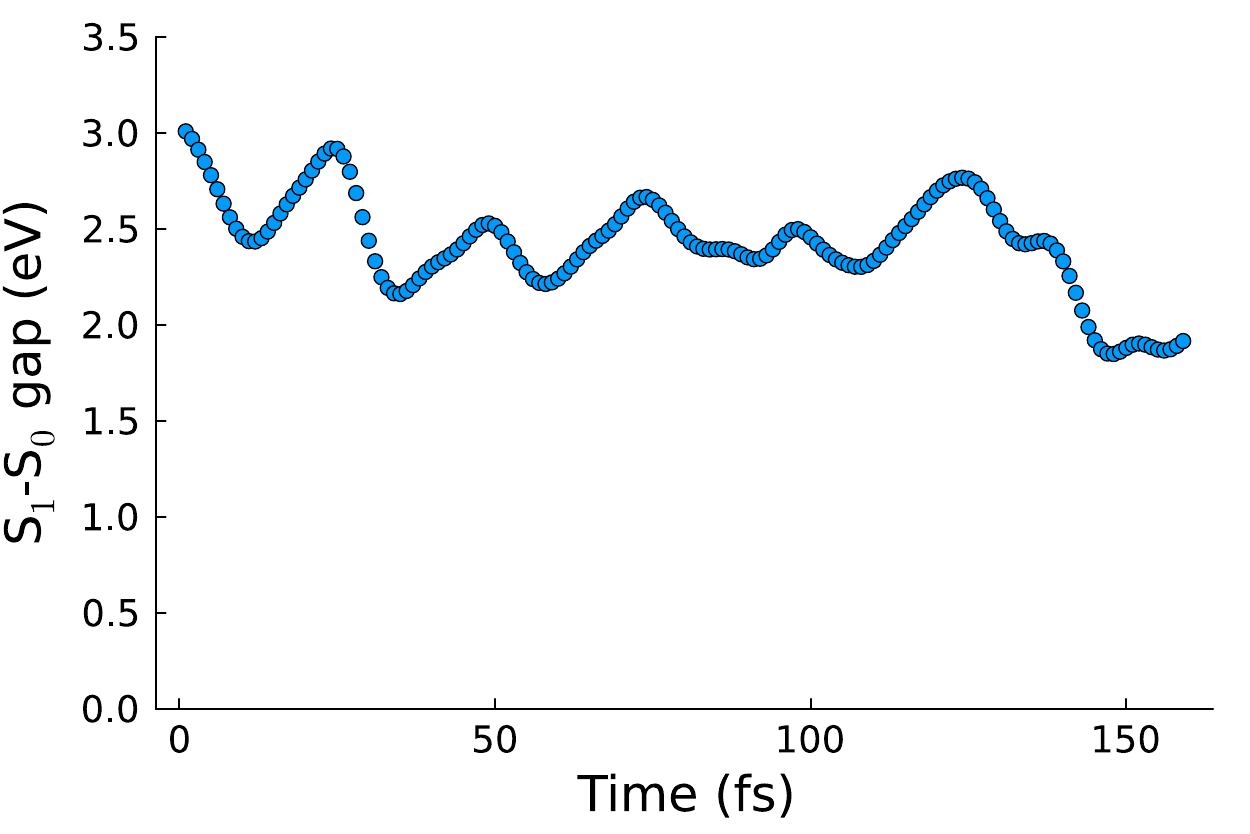}
     \end{subfigure}
        \caption{Energy gap between S$_0$ and S$_1$ for reactive dynamics at 1000 K of \textit{p}-DAPA-CN. The final time shown is the proton transfer time.}
        \label{fig_chap06:GapS2S1CN}
\end{figure}

\begin{figure}[bht!]
     \centering
     \begin{subfigure}[b]{0.3\textwidth}
         \centering
         \includegraphics[width=1.0\textwidth]{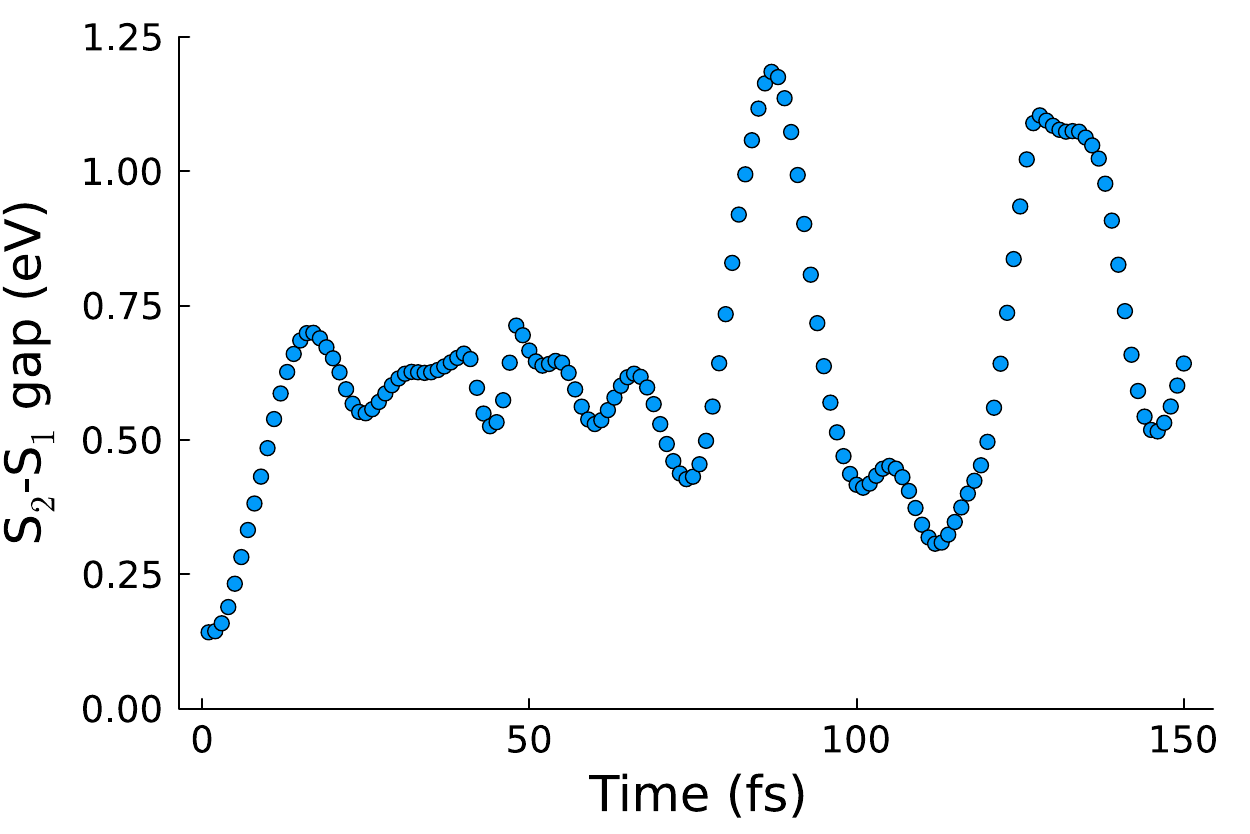}
     \end{subfigure}
     \hfill
     \begin{subfigure}[b]{0.3\textwidth}
         \centering
          \includegraphics[width=1.0\textwidth]{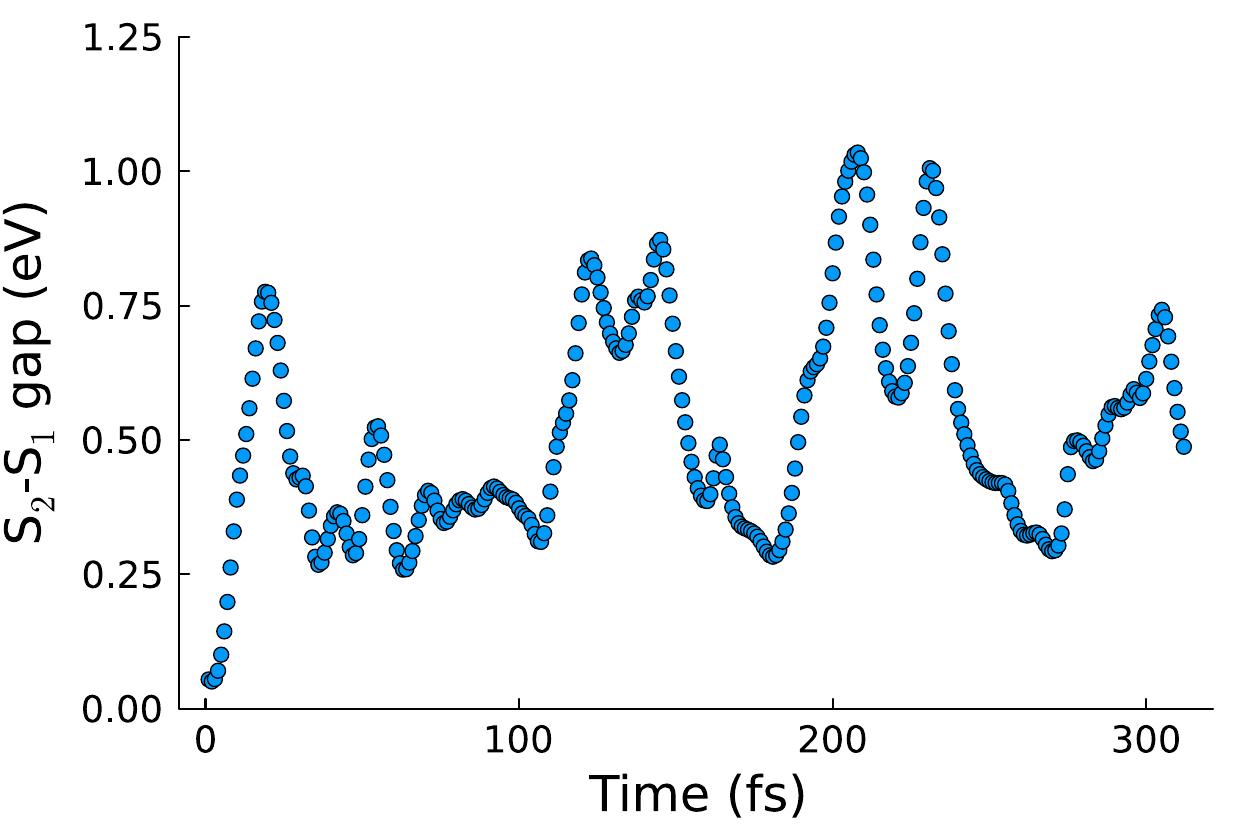}
     \end{subfigure}
     \hfill
        \begin{subfigure}[b]{0.3\textwidth}
         \centering
         \includegraphics[width=1.0\textwidth]{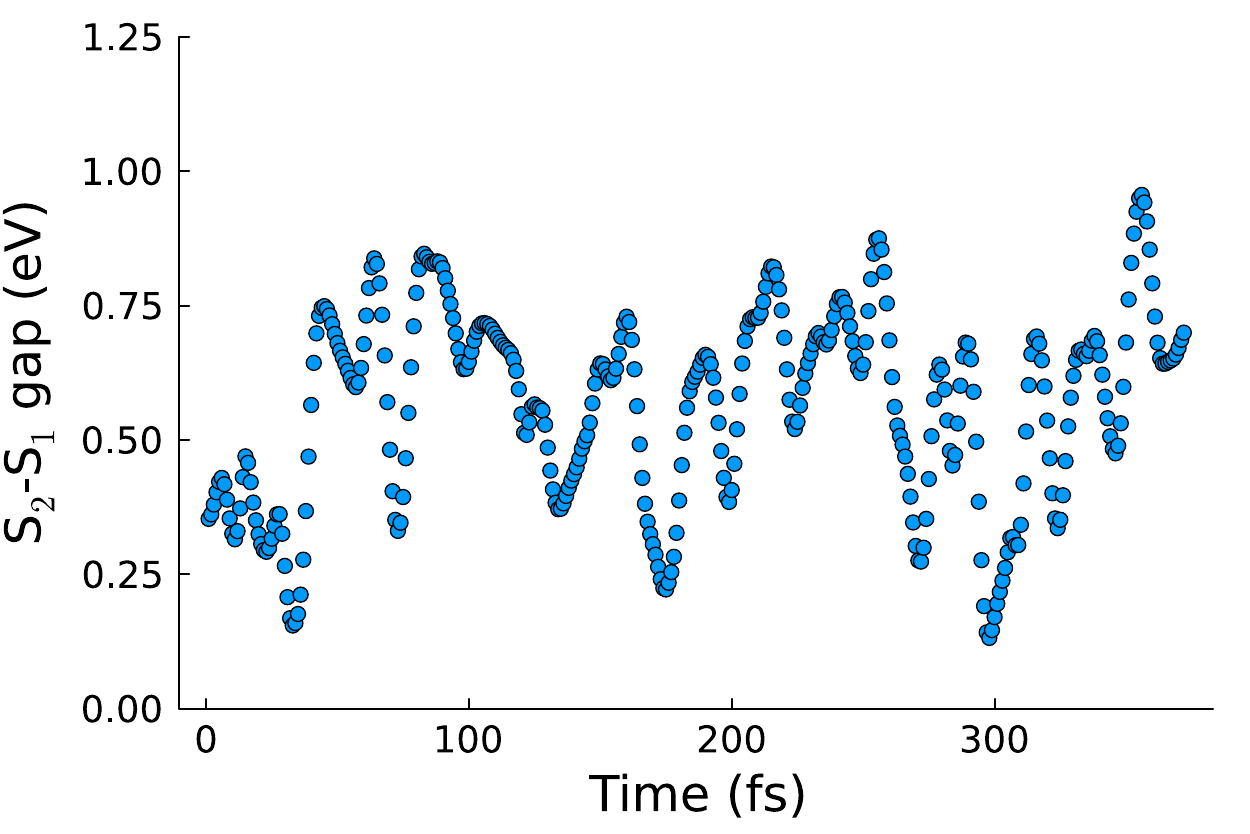}
     \end{subfigure}
     \hfill
        \begin{subfigure}[b]{0.3\textwidth}
         \centering
         \includegraphics[width=1.0\textwidth]{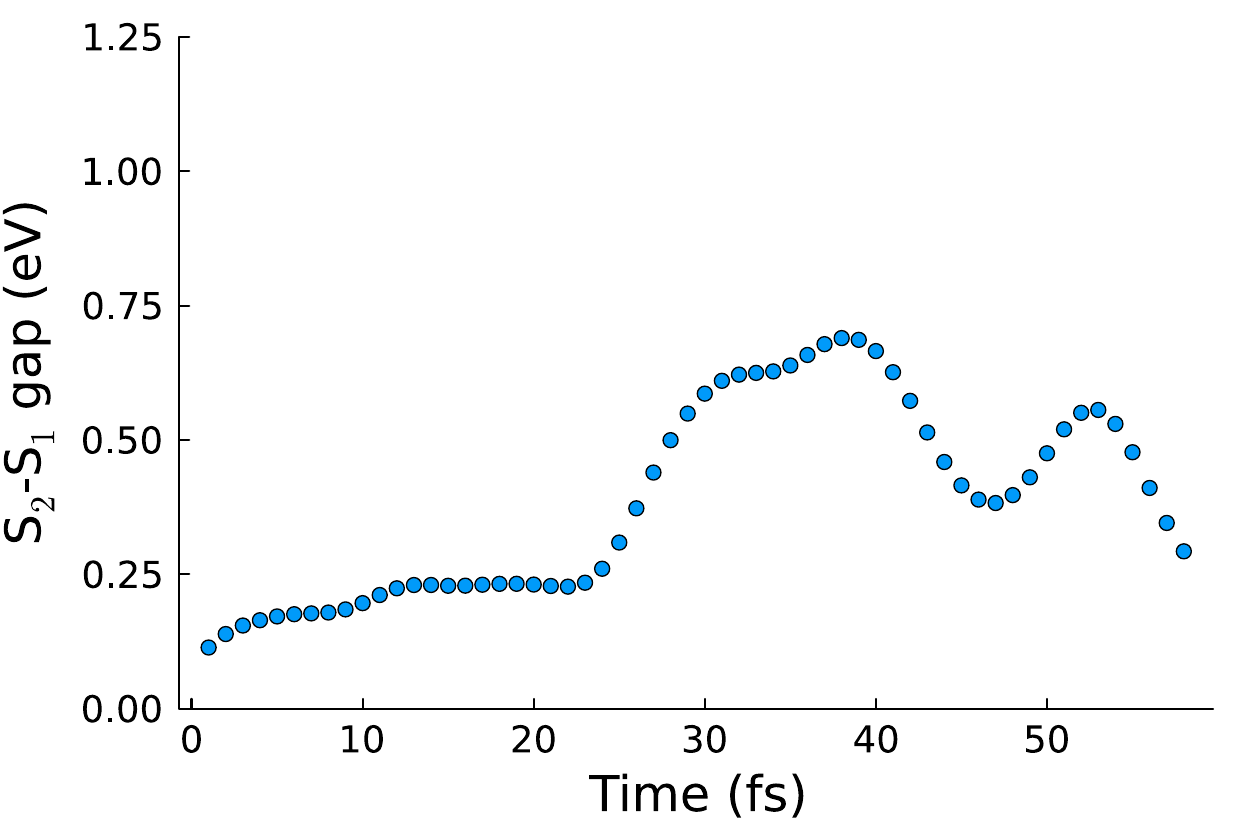}
     \end{subfigure}
     \hfill
        \begin{subfigure}[b]{0.3\textwidth}
         \centering
         \includegraphics[width=1.0\textwidth]{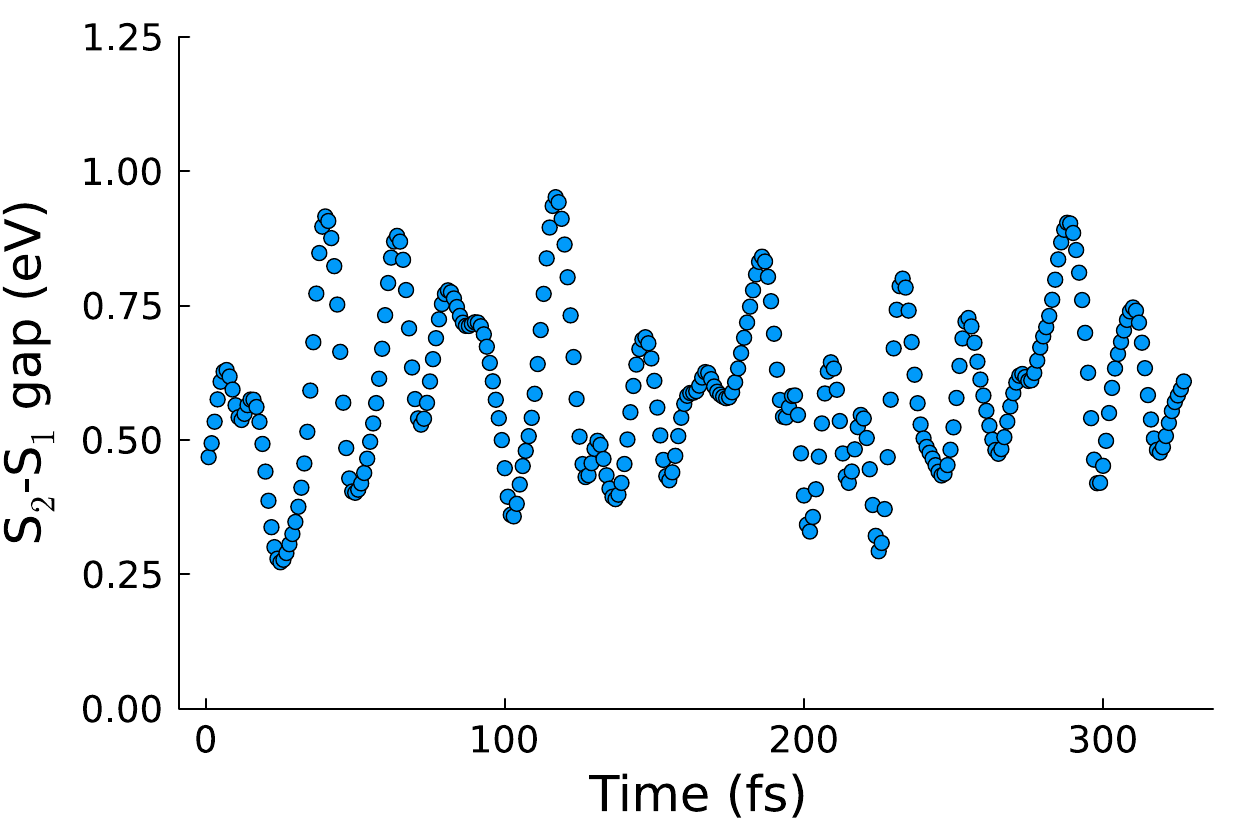}
     \end{subfigure}
        \hfill
        \begin{subfigure}[b]{0.3\textwidth}
         \centering
         \includegraphics[width=1.0\textwidth]{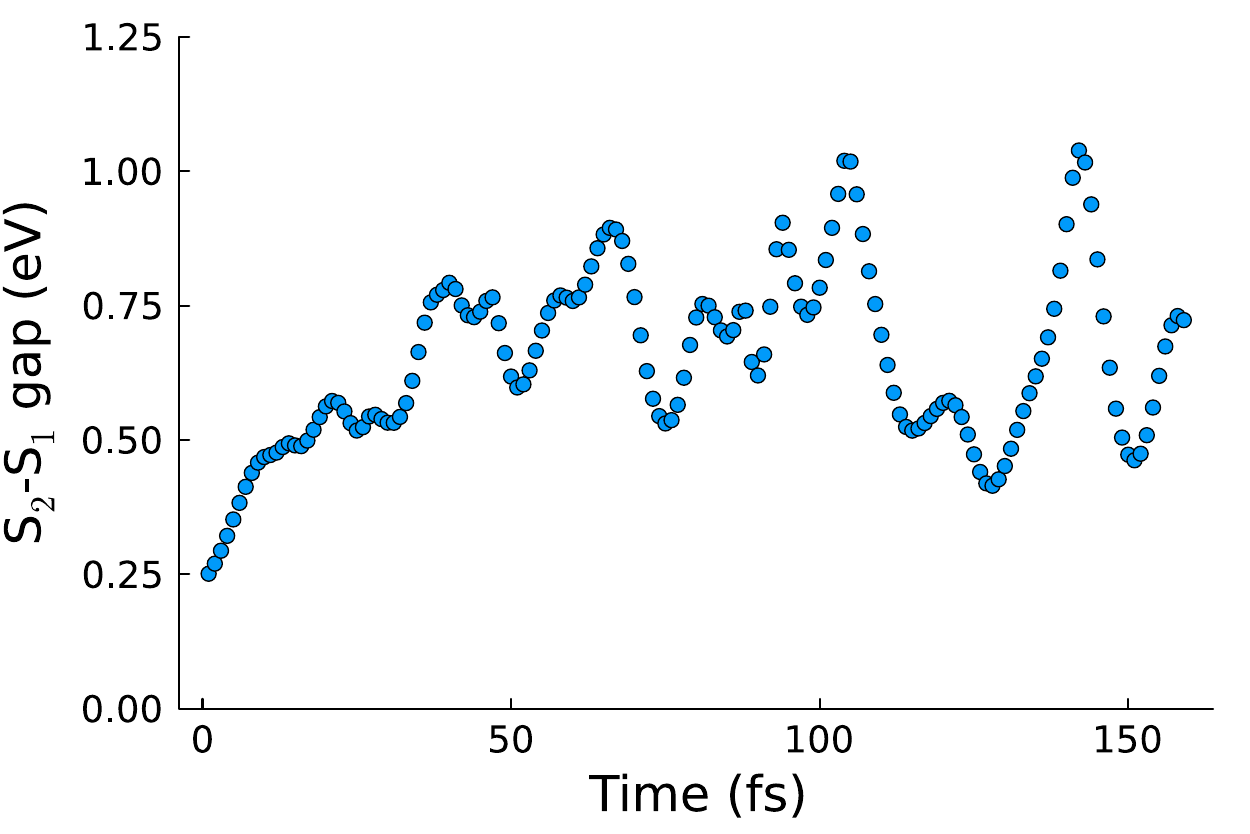}
     \end{subfigure}
        \caption{Energy gap between S$_2$ and S$_1$ for reactive dynamics at 1000 K of \textit{p}-DAPA-CN. The final time shown is the proton transfer time.}
        \label{fig_chap06:GapS2S1CN}
\end{figure}

\newpage
\,
\newpage

\subsection*{Vibrational Control at 300K}
We present in Table~\ref{table:vibcontrol_300K} the more detailed results of the dynamics under vibrational control at 300K. This table is used for the Table~2 in the main text.

 \begin{table}[ht]
 \centering
\caption{\textmd{Proton transfer time $\stau_\text{PT}$ in fs at 300 Kelvin where the four values by cells correspond to four different thermal sampling results. $\hbar \omega$ correspond to a one-photon absorption respective to the vibration frequency. \textcolor{red}{X} corresponds to no proton transfer whereas \textcolor{teal}{$\times 2$} corresponds to a double proton transfer.}}
\label{table:vibcontrol_300K}
\setlength{\aboverulesep}{0pt}
\setlength{\belowrulesep}{0pt}
\resizebox{\textwidth}{!}{%
\begin{tabular}{c |*{4}{c}| *{4}{c}| *{4}{c} |*{4}{c}}
& \multicolumn{4}{c|}{\makecell{$p$-DAPA-CF$_3$\\sym}} &  \multicolumn{4}{c|}{\makecell{$p$-DAPA-CF$_3$\\antisym}}  & \multicolumn{4}{c|}{\makecell{$p$-DAPA-CN\\sym}}  &  \multicolumn{4}{c}{\makecell{$p$-DAPA-CN\\antisym}} \\ 
\midrule 1.22 eV & 66 & 39 & 31 & 109    & 60 & 112 & 41 & \textcolor{red}{X}     & 71 & 71 & 36 \textcolor{teal}{($\times 2$)} & 70     & 45 & 49 & 14 & 16 \\
\midrule 0.620 eV & 456 & 106 & 35 & \textcolor{red}{X}    & 536 & 120 & 369 & \textcolor{red}{X}     & 188 & 102 & 37 & \textcolor{red}{X}     & 54 & 46 & 57 & 45 \\
\midrule 0.400 eV & \textcolor{red}{X} & 107 & 39 & \textcolor{red}{X}     & \textcolor{red}{X} & \textcolor{red}{X} & \textcolor{red}{X} & \textcolor{red}{X}     & \textcolor{red}{X} & \textcolor{red}{X} & 142 & 655     & 88 & 48 & 143 & 48   \\
\midrule 0.200 eV & \textcolor{red}{X}  &  108  & \textcolor{red}{X}  & \textcolor{red}{X}     & \textcolor{red}{X} & \textcolor{red}{X} & \textcolor{red}{X} & \textcolor{red}{X}     & \textcolor{red}{X} &  \textcolor{red}{X} & 141 &  \textcolor{red}{X}      &  \textcolor{red}{X} & 52 & 144 & 51  \\
\midrule $\hbar \omega$ & \textcolor{red}{X} & 110 & \textcolor{red}{X} & \textcolor{red}{X}        &  \textcolor{red}{X} & \textcolor{red}{X} & \textcolor{red}{X} & \textcolor{red}{X}           & \textcolor{red}{X} & \textcolor{red}{X} & 143 & \textcolor{red}{X}        & \textcolor{red}{X} & 60 & 145 & 55  \\
\midrule 0 eV & \textcolor{red}{X} & \textcolor{red}{X} & \textcolor{red}{X} & \textcolor{red}{X}        & \textcolor{red}{X} & \textcolor{red}{X} & \textcolor{red}{X} & \textcolor{red}{X}        & \textcolor{red}{X} & \textcolor{red}{X} & \textcolor{red}{X} & \textcolor{red}{X}        & \textcolor{red}{X} & \textcolor{red}{X} & \textcolor{red}{X} & \textcolor{red}{X} \\
\end{tabular}%
}

\end{table}

\subsection*{Analysis of the vibrational energy distribution}

\begin{figure}[bht!]
    \centering
    \hspace*{-1.5cm} \includegraphics[scale=0.28]{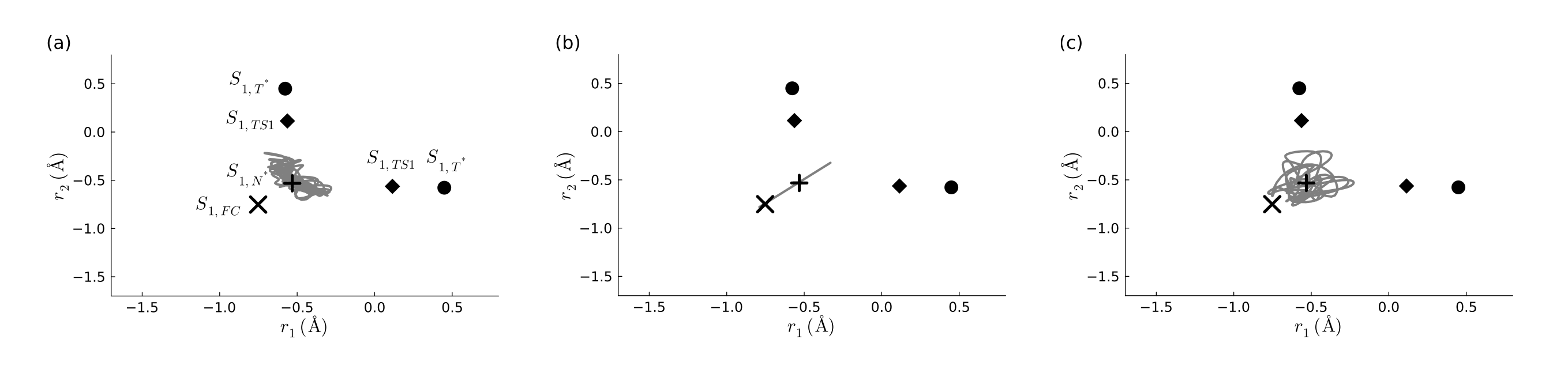}
    \caption{Dynamics at 0 K of $p$-DAPA-CF$_3$ for the excitation of (a) normal mode 51 at amplitude 2.48 eV, (b) normal mode 70 at amplitude 2.48 eV, (c) normal modes 51,53,54 and 70 simultaneously excited at amplitude 1.21 eV}
    \label{fig_sm:NM_notworking}
\end{figure}

\begin{figure}[bht!]
     \centering
                   \includegraphics[width=1.0\textwidth]{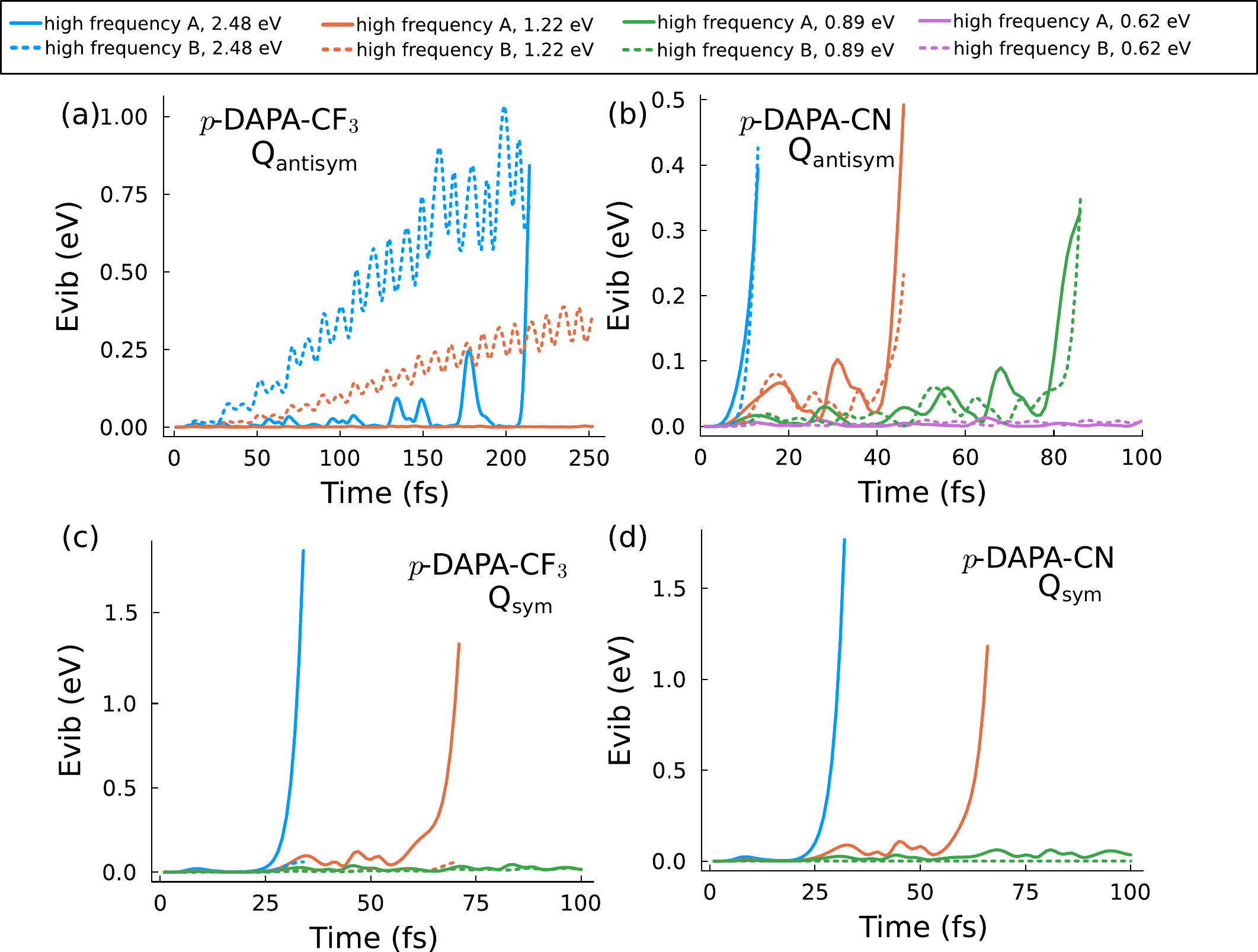}
          \caption{Vibrational energy over time injected into the high-frequency proton displacement modes (A for symmetric, B for antisymmetric) at 0K. The energy distribution for the different activation energy is shown (a) for $p$-DAPA-CF$_3$ with the activation of the antisymmetric identified mode, (b)  for $p$-DAPA-CN with the activation of the antisymmetric identified mode, (c) for $p$-DAPA-CF$_3$ with the activation of the symmetric identified mode, (d) for $p$-DAPA-CN with the activation of the symmetric identified mode.}
          \label{fig_sm:high_freq_energy}
\end{figure}

First, we show in figure~\ref{fig_sm:NM_notworking} some examples of  infrared activation on modes for $p$-DAPA-CF$_3$ that did not show a reaction at 0 Kelvin.
For reactive mode excitation at 0 Kelvin, we focus here on the vibrational energy distribution. In Fig.~\ref{fig_sm:high_freq_energy}, we show the excitation of high-frequency modes that cause the proton displacement towards the proton acceptor, without backbone motion (hydrogen stretching modes). A high energy in these modes is an indicator of a proton transfer in these examples, transcribing a change of the appropriate normal mode basis. In Fig.~\ref{fig_sm:high_freq_energy}~(c,d), the excitation of one high frequency modes of $p$-DAPA-CF3 and $p$-DAPA-CN follows the same trend for the symmetric activation. In Fig.~\ref{fig_sm:high_freq_energy}~(a,b), however, two different high-frequency modes are excited. For the case of $p$-DAPA-CN, the excitation is sudden whereas the increase is slower and more progressive for the $p$-DAPA-CF$_3$ case, illustrating a different energy increase pattern. That leads to a favourable, sudden combination for $p$-DAPA-CN and therefore a faster transfer compared to $p$-DAPA-CF$_3$. These energy distributions indicate a less straightforward energy path for $p$-DAPA-CF3 between the high frequency modes and the antisymmetric vibrationally activated mode in comparison with $p$-DAPA-CN. \\

\begin{figure}[bht!]
     \centering
                   \includegraphics[width=0.9\textwidth]{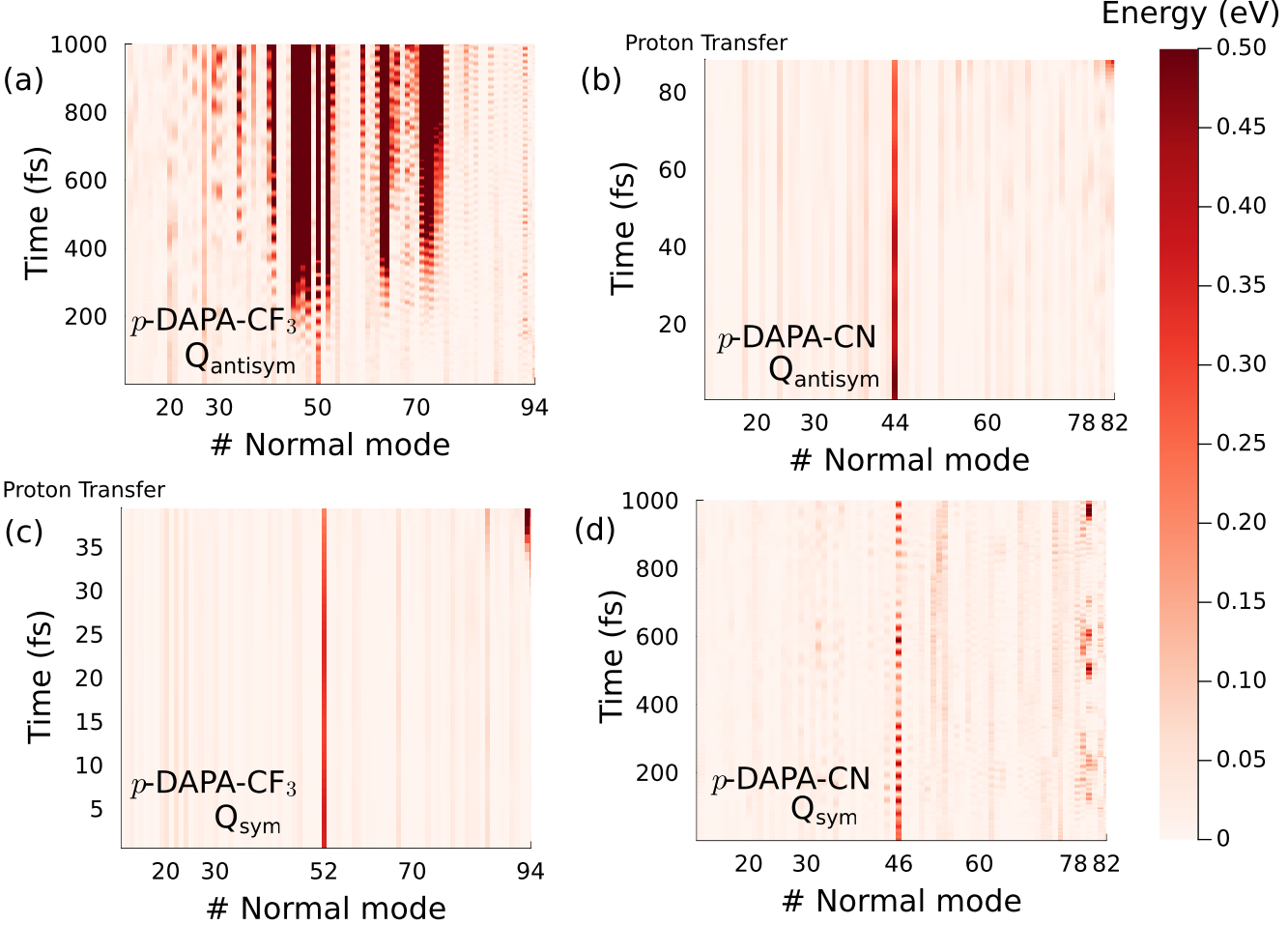}
          \caption{Vibrational energy distribution among normal modes over time at 300 K for (a) $p$-DAPA-CF$_3$  with an antisymmetric vibrational activation, (b) $p$-DAPA-CN  with an antisymmetric vibrational activation, (c) $p$-DAPA-CF$_3$  with a symmetric vibrational activation, (d) $p$-DAPA-CN  with a symmetric vibrational activation. For $p$-DAPA-CF$_3$, modes 50 and 52 are the activated mode and the modes 92 and 94 are the high frequency proton transfer modes. For $p$-DAPA-CN, the modes 46 and 44 are activated, the modes 82 and 81 are the high frequency proton transfer modes and the modes 78 and 79 are CN elongations. \\}
          \label{supmat:evibdistrib}
     \end{figure}

We show in Fig.~\ref{supmat:evibdistrib} the distribution  of vibrational energy among most of the normal modes over time done with the post-dynamics normal mode analysis. Each column corresponds to a normal mode in increasing frequency. For the non-reactive dynamics examples, it illustrates the broader distribution for $p$-DAPA-CF$_3$ and the transient excitation of CN elongations in $p$-DAPA-CN. 

\vspace*{-0.00cm}
\enlargethispage{1\baselineskip}

\newpage
\,
\newpage

\bibliography{SBF.bib}